\documentclass[a4]{article}
\usepackage[utf8x]{inputenc}
\usepackage{enumerate}
\usepackage{amsmath}
\usepackage{amssymb}
\usepackage{amsthm,amstext}
\usepackage{graphicx}
\usepackage{color}
\usepackage{subfigure}
\usepackage{psfrag}
\usepackage{auto-pst-pdf}
\usepackage{multirow}
\usepackage{mathtools}
\usepackage{empheq}
\usepackage[T1]{fontenc}
\usepackage{units}
\usepackage{color}
\newcommand{\Fabrice}[1]{\textcolor{black}{#1}}

\title{Asymptotic-Preserving methods and multiscale models for plasma physics}

\author{P. Degond$^\dagger$, F. Deluzet$^{\ddagger}$ \\[1em]
$^\dagger$Department of Mathematics, Imperial College London, \\
London SW7 2AZ, United Kingdom, \\
pdegond@imperial.ac.uk\\[1em]
$^\ddagger$Universit\'e de Toulouse; UPS, INSA, UT1, UTM,\\ Institut de Math\'ematiques de Toulouse,\\
CNRS, Institut de Math´ematiques de Toulouse UMR 5219,\\
F-31062 Toulouse, France,\\
fabrice.deluzet@math.univ-toulouse.fr \\[1em]
}

\date{}

\def\Rset{\mathbb{R}}

\newtheorem{remark}{Remark}[section]

\begin{document}
\maketitle
\begin{abstract}
The purpose of the present paper is to provide an overview of Asymptotic-Preserving methods for multiscale plasma simulations by addressing three singular perturbation problems. First, the quasi-neutral limit of fluid and kinetic models is investigated in the framework of non magnetized as well as magnetized plasmas. Second, the drift limit for fluid descriptions of thermal plasmas under large magnetic fields is addressed. Finally efficient numerical resolutions of anisotropic elliptic or diffusion equations arising in magnetized plasma simulation are reviewed.
\end{abstract}
\maketitle
\section{Introduction}
Plasma physics is by essence a multiscale problem \cite{bellan_fundamentals_2008,fitzpatrick_plasma_2014,huba_nrl_2011} mixing microscopic to macroscopic scales. The microscopic space scales describe the motion of particles, their collisions on the mean free path scale, and the interaction with the electromagnetic fields over the plasma skin depth or the Debye length and the Larmor radius. The macroscopic scales are characteristic of the field and the plasma macroscopic evolution. The diversity in the time scales is also very wide. They range from high frequency phenomena defined by the propagation of electromagnetic waves at the speed of light as well as the cyclotron and the plasma frequencies; to the evolution of the coupled system composed of the plasma and the fields on the time scales of the plasma mean flow evolution. The simulation of these systems on large scales have been the source of an intense and fruitful research, with the derivation, analysis and implementation of models relating different description levels, from the microscopic scales to mesoscopic and macroscopic descriptions \cite{bittencourt_fundamentals_2004,cercignani_many-particle_1997,cercignani_scaling_1998,Chen,krall_principles_1986,schnack_lectures_2009}.

Alongside, numerical methods have also been intensively developed with the same aim to produce simulations on large scales. Implicit methods \cite{Mas81,CLF82,HeLa87,jardin_review_2012,lapenta_particle_2012,toth_adaptive_2012,chen_multi-dimensional_2015} allow, to some extent, the de-correlation of the discretization parameters from the smallest scales described by the equations. The objective is thus to derive numerical methods with stability properties less restrictive than explicit methods. This gains the advantage of larger simulation parameters, chosen according to the scales of interest, rather than the small parameters described by the system. 

However implicit discretizations are ineffective when the stiffness of the problem leads to a degeneracy of the model. Fluid limits of kinetic equations are a good example of such frameworks. In the limit of a vanishing mean free path, the kinetic equation degenerates, constraining the distribution function to belong to the kernel of the collision operator. Unfortunately this only information does not permit to determine uniquely the distribution function. Therefore, the kinetic model becomes singular, while, in the limit of infinite collision rates, a fluid description is sufficient.
This difficulty is usually overcome thanks to domain decomposition strategies, consisting in using the asymptotic (macroscopic) description everywhere the multiscale (microscopic) model is not compulsory. However, this coupling strategy is still an open question, at least in specific contexts (see for instance \cite{jin_asymptotic_2012,pareschi_efficient_2011} and the references therein for examples of domain decomposition). The main difficulty raised by this strategy is twofold. First, it is not necessarily straightforward to match the unknowns advanced by the multiscale model with those of the limit one \cite{jin_asymptotic_2012}. Second, the interface delimiting the sub-domains may be evolving with the system, its location computation being a problem by its own. On top of that, this interface should be immersed in a part of the domain where both the multiscale and the limit models are valid. The existence of such a region supposes that the approximation error of the numerical method dominates the modelling error produced by the use of the limit model for non vanishing asymptotic parameter values. For refined numerical parameters this requirement may not be met (an example of such a situation is discussed in \cite{CDN14}), or at the price of important computational efforts.

An alternative approach has been introduced in the frame of Asymptotic-Preserving methods originally designed to cope with fluid and diffusive limits of kinetic equations \cite{jin_efficient_1999}. The purpose is to derive numerical methods bridging the microscopic description and the asymptotic model when this latter is derived as a singular limit of the former model. Fluid and diffusive limits of kinetic equations remain the most active field of developments for AP schemes outlined by numerous publications \cite{degond_smooth_2005,pareschi_implicitexplicit_2005,bennoune_uniformly_2008,liu_analysis_2010,crouseilles_asymptotic_2011,pareschi_efficient_2011,crestetto_kinetic/fluid_2012,dimarco_asymptotic_2012,mohammed_lemou_micro-macro_2012,dimarco_asymptotic_2012,boscarino_implicit-explicit_2013,carrillo_asymptotic_2013,jang_high_2015,james_numerical_2015,CDV,lafitte_high-order_2016}. Other singular limits are however addressed, among which the low Mach regime for fluid systems \cite{degond_mach-number_2007,degond_all_2011,haack_all-speed_2012,cordier_asymptotic-preserving_2012,tang_second_2012,noelle_weakly_2014} with applications to semiconductors \cite{dimarco_implicit-explicit_2013,bessemoulin-chatard_study_2014}, nano-structures and quantum systems \cite{degond_time_2008,besse_asymptotic_2013,crouseilles_asymptotic_2014,hu_asymptotic-preserving_2014} this list being non exhaustive. We also refer to \cite{jin_asymptotic_2012,degond_numerical_2013} for reviews on AP methods.

AP-methods are in general implicit schemes. However, they differ from classical implicit discretizations in their derivation. Indeed, AP-methods are constructed to bridge several sets of equations describing the system in different regimes, rather than alleviating the stability constraints of the numerical methods. The aim here is to shape a single set of reformulated equations unifying these regimes. Asymptotic-Preserving schemes offer a consistent discretization of the multiscale model when the discretization parameters resolve the small scales. Conversely, the AP-methods are consistent with the limit model when the numerical parameters are large compared to the microscopic scales. In this reformulated system the limit is regular, which means that the macroscopic problem is recovered for vanishing asymptotic parameters. Three main steps can be identified in the derivation of these methods. The first one consists in elaborating the limit problem. This system is defined as the set of well posed equations providing the limit of the solution in the asymptotic regime. In the framework of singular perturbation problems, the limit problem is not readily obtained by setting the asymptotic parameter to zero in the multiscale problem. The second step consists in deriving the reformulated set of equations mentioned above. In this aim a good understanding of the derivation of the limit problem is mandatory. Finally the time discretization is designed to meet consistency requirements with all regimes. This necessitates a sufficient implicitation level in order to guaranty the consistency with the different regimes, providing thus a means of computing all the unknowns whatever the values of the asymptotic parameter. 

The purpose of the present paper is to illustrate these concepts thanks to three singular perturbation problems.

The first one is related to the quasi-neutral limit of fluid and kinetic plasma models coupled to the Maxwell system. In many applications the charge separations can be disregarded at the scales of interest, the plasma being thus considered as quasi-neutral. This property is harnessed to derive reduced models, filtering the scales describing the charge separations out from the equations. However, an accurate description of the whole system on large scales generally requires the resolution of the complex physics occurring in limited regions, for instance near the plasma boundary or, to describe the interaction of the plasma with a wall (see \cite{langmuir_interaction_1929,post_physics_1986,stangeby_plasma_2000}) where electrostatic sheathes may appear. In these regions the assumptions used to derive quasi-neutral models are not valid. More complex models are needed to account for the physics prevailing in these regions, calling for numerical methods able to treat efficiently quasi-neutral models with local break down. The difficulty of this asymptotic is explained by the degeneracy of the equation providing the electric field. In non quasi-neutral models, the electric field is computed from the Maxwell system \cite{BiLa04,HoEa88,taflove_computational_2005}. In the quasi-neutral regime, these equations being degenerate, the electric field is computed from the equations describing the evolution of the electrons and the ions. These regimes translate thus two different physical phenomena, accounting for the singular nature of this limit addressed in Sec.~\ref{sec:QN}.

The second context is specific to hot plasmas evolving in the presence of strong magnetic field. These investigations are limited to fluid descriptions with an external magnetic field. The purpose here, is to address the so called gyro-fluid regime \cite{hammett_developments_1993} which bears some analogies with drift approximations \cite{biskamp_nonlinear_1997,krall_principles_1986,spatschek_high_2012,schnack_lectures_2009,grandgirard_drift-kinetic_2006,tamain_tokam-3d:_2010,ricci_simulation_2012,littlejohn_variational_1983} largely used for the modelling of tokamak plasmas on large time scales. The drift regime is characterized by a vanishing inertia, which is responsible of the momentum equation degeneracy. Along the magnetic field lines, this equation imposes a zero force regime with the pressure gradient balancing the electric force. No explicit occurrence of the parallel momentum survives in the drift asymptotic which explains the singularity of this limit. A regimes transition is observed near the wall, with a significant decrease of the plasma temperature and more importantly an acceleration of the particles by the electrostatic field existing in the sheath (\cite{chodura_plasma_1986,stangeby_plasma_2000}). In this area the particle inertia becomes again significant breaking the drift assumption. The derivation of asymptotic preserving methods in this framework is detailed in Sec.~\ref{sec:drift}.

Finally, we propose a review of AP-methods for anisotropic elliptic or diffusion equations arising in the simulation of tokamak \cite{crouseilles_comparison_2015} and ionospheric  \cite{besse_model_2004} plasmas. In a plasma under large magnetic fields, the diffusion along the magnetic field lines is almost infinite in the time scale of the dynamic in the transverse directions \cite{gunter_modelling_2005,sharma_preserving_2007,del-castillo-negrete_parallel_2012,filbet_numerical_2012}. The difficulty raised by this anisotropy in the context of tokamaks is due to the periodicity of the torus. Indeed, in the limit of an infinite diffusion in the aligned direction, the parallel diffusion is dominant, this operator being supplemented with periodic boundary conditions at each ends of the magnetic field lines. Therefore, its kernel is not reduced to zero, leading to a difficulty comparable to the one referred above concerning the limit of infinite collisions for kinetic descriptions. This issue is illustrated thanks to a simplified problem in the beginning of Sec.~\ref{sec:aniso} and investigated by means of AP-methods in the sequel of the section.

\section{Quasi-neutral limit of kinetic and fluid plasma descriptions}
\label{sec:QN}
\subsection{Introduction}
\label{sec:QN:intro}

The quasi-neutral limit can be related to the so-called {\it plasma approximation} as defined in \cite{Chen} which consists, for dense plasmas, in assuming equal ionic and electronic densities $n_i = n_e$ together with an electric field that is not divergence free $\nabla \cdot E \neq 0$. This might appear as a paradox since it breaks the Maxwell-Gauss equation $\nabla \cdot E = q(n_i - n_e)/\varepsilon_0$ ($\varepsilon_0$ being the vacuum permittivity and $q$ the elementary charge).

\Fabrice{This ambiguity is clarified in Sec.~\ref{sec:QN:analysis} thanks to the analysis of the different orderings revealed by a scaling of the Maxwell system. 
The inter-relations of the dimensionless parameters occurring in this set of re-scaled equations define different asymptotic regimes. The first one relates the propagation of an electromagnetic wave at the speed of light and is referred to as the Maxwell regime. In this ordering, the Maxwell sources vanish and do not contribute to the changes in the electromagnetic field, the electric field being computed by means of the displacement current in Amp\`ere's law.
On the contrary, the evolution in the quasi-neutral regime is dominated by the sources with a negligible displacement current. Therefore, the transitions between the Maxwell and the quasi-neutral regimes rely on the relative influence of the displacement current and the current of particles. The vanishing of the displacement current in the Amp\`ere equation is at the origin of the singular nature of the quasi-neutral limit.
Finally the electrostatic limit of the Maxwell system is discussed, the aim being to characterize the quasi-neutral limit in this asymptotic. It is defined as a low frequency regime with a slow system evolution compared to the speed of light. In particular, the singular nature of the quasi-neutral limit is illustrated also in this framework, with the degeneracy of Maxwell-Gauss equation.}

\Fabrice{In the quasi-neutral asymptotic, the degeneracy of the Amp\`ere and the Gauss equations call for new means of computing the electric field. This is classically achieved thanks to a generalized Ohm law, either through the quasi-neutral constraint enforcing a divergence free current of particles or, the electronic momentum equation, in which the inertia is neglected. We refer to \cite{langmuir_interaction_1929} for some seminal works on quasi-neutral models, to \cite{hewett_low-frequency_1994} for a short review, \cite{winske_hybrid_1991,joyce_electrostatic_1997,buchner_hybrid_2003,crispel_quasi-neutral_2005,crispel_plasma_2007,tronci_hybrid_2014,tronci_neutral_2015} and the references therein for implementations of quasi-neutral plasma models.}

\Fabrice{However quasi-neutral descriptions have a limited range of validity. In particular these models are not valid in vacuum or low plasma density regions where high frequency phenomena may occur \cite{tonks_general_1929}. 
The purpose of the Asymptotic-Preserving methods reviewed in this paper, is to bring the two regimes into a single set of equations, making possible the transition between the evolution of the electric field in the Maxwell regime, by means of the displacement current and, that of the quasi-neutral regime, with an electric field computed thanks to a generalized Ohm law. In this respect, AP methods implement the guide line stated in \cite{Chen} ``{\it do not use Maxwell's equations to compute the electric field unless it is unavoidable !}''.}

\Fabrice{The derivation of the AP methods is addressed in Sec.~\ref{sec:AP:VM}. One key point is to identify the equations describing the system in the quasi-neutral regime. The Ohm's law heavily relying on the equations describing the particles evolution, the plasma models, either fluid or kinetic, are therefore an important aspect to consider in designing AP methods. Pioneering works have been first devoted to the Euler-Poisson system \cite{CDV07}, then extended to kinetic electrostatic descriptions by means of the Vlasov-Poisson system \cite{degond_asymptotically_2006,degond_asymptotic-preserving_2010,BCDS09}. Electromagnetic fields have been considered in the frame of the bi-fluid isothermal Euler-Maxwell system in \cite{degond_numerical_2012} (extended to the M1-Maxwell model in \cite{guisset_asymptotic-preserving_2016}) and finally with the Vlasov-Maxwell system \cite{DegDelDoy}. In Sec.~\ref{sec:AP:VM} a unified presentation of the different regimes is proposed by means of the ``augmented'' Vlasov-Maxwell system. This choice offers different advantages. First, the augmented system contains the difficulty of both the electromagnetic and the electrostatic regimes. Second, the Ohm law being derived from equations driving the evolution of macroscopic quantities, the construction of AP methods can be readily transposed from the kinetic to the fluid framework. Finally an overview of the numerical implementations of AP-methods is proposed in Sec.~\ref{sec:QN:num}.}

\subsection{Outlines of the quasi-neutral and electrostatic limits of the Maxwell system}\label{sec:QN:analysis}
\Fabrice{The objective of this section are twofold. On the one hand, the quasi-neutral limit is investigated  thanks to a scaling of the equations in order to explain the paradox raised in the introduction. On the other hand, some consistency properties, relating the Maxwell-Amp\`ere and the Maxwell-Gauss equations are outlined. The electrostatic limit of the Maxwell system is also discussed. In this aim, the Maxwell system is complemented with the continuity equation driving the evolution of the charge density, defining the system of interest}
\begin{subequations}\label{eq:Maxwell:sys:dim}
\begin{align}
& \frac{1}{c^2} \frac{\partial E}{\partial t} -  \nabla \times B=- \mu_0 J\,,\label{M-Ad}\\
& \frac{\partial B}{\partial t} + \nabla \times E = 0\,,\label{M-Fd}\\
& \nabla \cdot E= \frac{\rho}{\varepsilon_0}\,,\label{M-Gd}\\
& \nabla \cdot B = 0\,,\label{M-Td}
\end{align}  
\end{subequations}
\begin{equation}
  \frac{\partial \rho}{\partial t} + \nabla \cdot J = 0\,, \label{Cd}
\end{equation}
consisting of the Maxwell-Amp\`ere \eqref{M-Ad}, the Maxwell-Faraday \eqref{M-Fd}, Maxwell-Gauss \eqref{M-Gd} and the Maxwell-Thomson \eqref{M-Td}
 equations supplemented with the continuity equation \eqref{Cd}.
In these equations $(E,B)$ is the electromagnetic field, the charge and current densities are defined by the electronic and ionic densities and mean velocities as $\rho=q(n_i-n_e)$ and $J=q(n_iu_i-n_eu_e)$, $q$ denoting the elementary charge. Finally $c$ is the speed of light, $\mu_0$ and $\varepsilon_0$ being the vacuum permeability and permittivity, verifying $\varepsilon_0 \mu_0 c^2=1$. 

\Fabrice{The physical variables are scaled by their typical value: $\bar x$ and $\bar t$ being the space and time scales, the following identity holds $x=\bar x \, x'$ and $t=\bar t \, t'$, $x'$ and $t'$ denoting the dimensionless variables. These scales define  $\bar \vartheta = \bar x / \bar t$ the velocity driving the changes in the electromagnetic field, the typical magnitude of the electric and magnetic fields being $\bar E$ and $\bar B$. The plasma characteristics are denoted $\bar T$, $\bar u$ and $\bar n$  for the typical temperature, mean velocity and density, allowing the definition of  $\lambda_D = (\varepsilon_0 k_B \bar T / (q^2 \bar n))^{1/2}$ the Debye length, with $k_B$ the Boltzmann constant and the electronic thermal velocity $v_{th,e}=(k_B \bar T /m_e)^{1/2}$, with $m_e$ the electronic mass. The Maxwell sources are scaled with $\bar \rho =q \bar n $ and $\bar J = q \bar n \bar u$ for the charge and current densities.
The introduction of the re-scaled variables into the equations reveals some dimensionless parameters:
\begin{equation}\label{eq:dimensionless:parameters}
  \begin{cases}
 \displaystyle \alpha=\frac{\bar \vartheta}{c}, \text{ the typical velocity to the speed of light}\,;\\[3mm]
 \displaystyle \zeta=\frac{\bar u}{ \bar \vartheta}, \text{ the plasma mean velocity relative to the speed of interest}\,; \\[3mm]
 \displaystyle M= \frac{\bar u}{v_{th,e}}\, \text{ the electronic Mach number, } v_{th,e}^2=\frac{k_B \bar T}{m_e}\,;\\[3mm]
 \displaystyle \eta = \frac{q \bar E \bar x}{m_e \bar u^2}, \text{ the ratio of the electric and plasma kinetic energies}\,; \\[3mm]
 \displaystyle \beta=\frac{\bar \vartheta \bar B}{\bar E},  \text{ the induced electric field to the total electric field}\,;\\[3mm]
  \displaystyle \lambda = \frac{\lambda_D}{\bar x} , \text{ the dimensionless Debye length, } \lambda_D^2 = \frac{\varepsilon_0 k_B \bar T}{q^2 \bar n} \,.
  \end{cases}
 \end{equation}}
 With these dimensionless parameters, the scaled system is recast into
\begin{align}
& \lambda^2 \frac{\partial E}{\partial t} -  \beta\frac{\lambda^2}{\alpha^2} \nabla \times B=- \frac{\zeta}{\eta M^2} J\,,\tag{M-A}\label{M-A}\\
& \beta \frac{\partial B}{\partial t} + \nabla \times E = 0\,,\tag{M-F}\label{M-F}\\
&\lambda^2 \eta M^2 \nabla \cdot E= \rho\,,\tag{M-G}\label{M-G}\\
& \nabla \cdot B = 0\,,\tag{M-T}\label{M-T}\\[2mm]
& \frac{\partial \rho}{\partial t} +  \zeta \nabla \cdot J = 0\,, \tag{C}\label{C}
\end{align}
where, for sake of readability, the primes are omitted for the scaled variables.

Two regimes can be identified accordingly to the frequency range characterizing the system evolution. In the high frequency limit, referred to as the Maxwell regime in the sequel, the velocity of interest is assumed comparable to the speed of light and large compared to both the mean velocity of the plasma and the particles thermal velocity. The Debye length is assumed to be large or comparable to the typical space scale. This translates into the following scaling relations
\begin{equation}\label{eq:regime:Maxwell}
\lambda^2=\alpha^2 = \beta = \eta = M^2 =  1\,,  \quad \zeta \ll 1 \,.
\end{equation}
In the Maxwell regime the plasma evolution can be disregarded. In the low density approximation $\lambda\gg 1$, the system reduces to the homogeneous Maxwell equations with the propagation of the electromagnetic wave at the speed of light. The electric field is computed by means of the displacement current in \eqref{M-A}.

The quasi-neutral limit is defined by a speed of interest comparable to the plasma mean velocity and the particles thermal velocity, these velocities being assumed small compared to the speed of light. The regime is therefore a low frequency asymptotic. The scaled Debye length is also assumed to define a small scale in this regime which finally yields to the scaling relations 
\begin{equation}\label{eq:regime:QN}
  \zeta=\beta = \eta = M^2 =  1\,,  \quad \lambda^2=\alpha^2 \ll 1 \,.
\end{equation}
The assumption $\beta = 1$ is common to all MHD models and referred to as the frozen field assumption. It translates that, in a dense plasma, the magnetic field is convected with the plasma flow. In particular, the propagation of electromagnetic waves at the speed of light is not possible in a dense plasma \cite{krall_principles_1986,Chen,bittencourt_fundamentals_2004,fitzpatrick_plasma_2014} and therefore not described by quasi-neutral models. The other dimensionless parameters are assumed equal to one in order to simplify the writing.
The Maxwell-Gauss equation \eqref{M-G} provides the quasi-neutrality $\rho = 0$, the electric field contribution vanishing in this equation. Therefore, this allows for the computation of non divergence free electric field and explains the paradox raised by a crude review of the dimensional equations. The Amp\`ere equation also degenerates in the asymptotic $\lambda^2 \to 0$ along with $\alpha^2/\lambda^2=\mathcal{O}(1)$, with a vanishing displacement current, the equation \eqref{M-A} yielding  
\begin{equation*}
  \nabla \times B = J \,.
\end{equation*}
In this equation too, the occurrence of the electric field vanishes in the quasi-neutral regime which leads to the conclusion that the (homogeneous) Maxwell equations cannot be used to compute the electric field in this limit. More precisely, the electric field must be found from the particles equation of motion \cite{Chen} by means of a generalized Ohm's law, explaining how the current of particles $J$ and the electric field relate to each other. This is routinely implemented in quasi-neutral descriptions of plasmas, the most widely used being the Magneto-Hydro-Dynamic (MHD) models \cite{biskamp_nonlinear_1997,davidson_introduction_2001,freidberg_ideal_2014,spatschek_high_2012}. 

Different frameworks have been investigated in this direction. The first one is devoted to electrostatic descriptions that can be derived from the dimensionless system above by letting $\alpha \to 0 $. Indeed, by the equation \eqref{M-A} the magnetic field is curl free ($\nabla \times B=0$), which together with $\nabla \cdot B = 0$ and assuming adequate boundary conditions, provides a constant magnetic field. The equation \eqref{M-F} provides thus a curl free electric field $\nabla \times E = 0$ assumed to derive from a potential $E=-\nabla \phi$. In this regime, the Maxwell-Amp\`ere equation can be decomposed into a curl free and a divergence free identity
\begin{equation}\label{eq:def:L:T:decomp}
  \begin{split}
   \lambda^2 \frac{\partial}{\partial t} \Delta \phi &= \frac{\zeta}{\eta M^2} J_L \,, \\
   \beta \frac{\lambda^2}{\alpha^2} \nabla \times B &= \frac{\zeta}{\eta M^2} J_T\,,    
  \end{split}
\end{equation}
where $J=J_L + J_T$, $\nabla \cdot J_T = 0$ and $ \nabla \times J_L =0 $. While $\nabla \times B $ vanishes in the electrostatic regime $\alpha\to0$, the quantity $\nabla \times B / \alpha^2$ remains finite \Fabrice{as long as the current transverse part $J_T$ does not vanish}. However it does not contribute to the definition of the electrostatic field, as outlined by the decomposition \eqref{eq:def:L:T:decomp}. This feature can also be recovered by computing formally the divergence of Amp\`ere's law \eqref{M-A}, providing
\begin{equation*}
  \lambda^2 \frac{\partial}{\partial t} \Delta \phi = \frac{\zeta}{\eta M^2}\nabla \cdot J \,.
\end{equation*}
This equation together with the continuity equation \eqref{C} provides
\begin{equation*}
  - \lambda^2 \eta M^2 \frac{\partial} {\partial t} \Delta \phi = \frac{\partial \rho}{\partial t}\,.
\end{equation*}
This outlines that the Gauss law is a consequence of the Amp\`ere \eqref{M-A} and the continuity \eqref{C} equations. The consistency of the initial condition with the Maxwell-Gauss equation \eqref{M-G} is preserved with time. Consequently, in the electrostatic regime, the only Maxwell-Gauss equation is sufficient to compute the entire electric field, this equation being usually substituted to the whole Maxwell system in this regime. Note that, the quasi-neutral limit is thus defined by $\lambda^2\to 0$ together with $\lambda^2/\alpha^2\to 0$, in contrast to $\lambda^2\to 0$ and $\lambda^2/\alpha^2= \mathcal{O}(1)$ for the electromagnetic framework.

\subsection{Asymptotic-Preserving formulation of the Vlasov-Maxwell system}\label{sec:AP:VM}
\subsubsection{The scaled Vlasov-Maxwell system}
The model investigated here consists of the Maxwell system \eqref{eq:Maxwell:sys:dim} coupled to a Vlasov equation for the electrons, the ions being assumed at rest with a uniform density to simplify the notations. The distribution function, denoted $f$, depends on  $x\in \Omega_x \subset  \Rset^3$, the microscopic velocity $v \in \Omega_v \subset \Rset^3$ and on time $t \in \Rset^+$. The function is the solution to
\begin{equation}
   \frac{\partial f}{\partial t} +  v \cdot \nabla f - \frac{q}{m_e}   ({E}+ v \times B)\cdot \nabla_v f =0
\end{equation}
 In order to address straightforwardly the asymptotic regime, the scaling defined by Eq.\eqref{eq:dimensionless:parameters} is again harnessed, but ti simplify, with $\bar u=\bar \vartheta$, where $\bar \vartheta=\bar x/\bar t$ and $\bar u$ is the mean plasma velocity, which amounts to setting  $\zeta=1$. The particles velocity $v$ being scaled  with the electronic thermal velocity $v_{th,e}=(k_B \bar T/m_e)^{1/2}$. 
In the sequel similar scaling relations as the ones defining the quasi-neutral regime \eqref{eq:regime:QN} will be considered. To simplify further the writing, the two small scales $\alpha^2$  and $\lambda^2$ will be denoted by a single parameter ($\lambda^2$), so that the quasi-neutral regime is easily identified by the limit $\lambda^2\to0$. The dimensionless Vlasov-Maxwell system is
\begin{subequations}\label{VM:scaled} 
 \begin{empheq}[left=(VM)^\lambda \empheqlbrace]{align}
&\frac{\partial f}{\partial t} +  v \cdot \nabla f -   ({E}+ v \times B)\cdot \nabla_v f =0\label{VM:scaled:V}\\
& \lambda^2  \frac{\partial {E}}{\partial t} -  \nabla \times B = -J,\label{VM:scaled:A}\\
&  \frac{\partial B}{\partial t} + \nabla \times E = 0,\label{VM:scaled:F}\\
&\lambda^2  \nabla \cdot {E}= 1 - n, \label{VM:scaled:G} \\
& \nabla \cdot B = 0\,, 
 \end{empheq}  
with 
\begin{equation}
  n=\int_{\Omega_v}f(x,v,t)\ dv\,, \qquad J=- \int_{\Omega_v}f(x,v,t)v\ dv\,.
\end{equation}
\end{subequations}
The sources of the Maxwell system verify a continuity equation derived from the moments of the Vlasov equation \eqref{VM:scaled:V} giving rise to
\begin{subequations}\label{eq:Moments}
  \begin{align}
    & \frac{\partial n}{\partial t} - \nabla \cdot  J = 0 \,, \label{eq:moment:0}\\ 
    & \frac{\partial J}{\partial t} - \nabla \cdot \mathbb{S} =  (n E - J \times B)\,, \qquad \mathbb{S}=\int_{\Omega_v}f(x,v,t)v \otimes v\ dv \,. \label{eq:moment:1}
  \end{align}
\end{subequations}

As outlined in section~\ref{sec:QN:analysis}, the Gauss equation is a consequence of the Amp\`ere law \eqref{VM:scaled:A} and the continuity equation \eqref{eq:moment:0}. However, the consistency with this latter is not always satisfied by numerical methods. This is for instance a common flaw of Particle-In-Cell methods largely documented (see for instance \cite{BiLa04,BCS07}). The most widely adopted solution is the correction of the electric field predicted by the Amp\`ere equation. This correction is computed by an electrostatic potential $p$ verifying the Maxwell-Gauss equation \eqref{VM:scaled:G}. This is the so-called Boris correction \cite{boris_proceedings_1970} decomposed in two steps. First the predicted electric field $\tilde E$  is computed by means of Amp\`ere's law. Second the correction is applied to this field, defining the corrected field $E = \tilde{E} - \nabla p$ in order for the Maxwell-Gauss equation to be satisfied:
\begin{subequations}
\begin{equation}
\lambda^2  \nabla \cdot {E}= 1 - n\,, \qquad  E = \tilde{E} - \nabla p \,, \label{eq:aVM:Corr:0} \qquad 
\end{equation}  
\end{subequations}
This gives rise to the dimensionless Vlasov-Maxwell system augmented with the corrector $p$
 \begin{subequations}\label{eq:aVM} 
 \begin{empheq}[left=(aVM)^\lambda \empheqlbrace]{align}
&\frac{\partial f}{\partial t} + v \cdot \nabla f - ({E}+ v \times B)\cdot \nabla_v f =0\label{eq:aVM:V}\\
& \lambda^2  \frac{\partial \tilde{E}}{\partial t} -  \nabla \times B = - J,\label{eq:aVM:A}\\
&  \frac{\partial B}{\partial t} + \nabla \times \tilde{E} = 0,\label{eq:aVM:F}\\
&\lambda^2  \Delta p = \lambda^2 \nabla\cdot \tilde{E} - (1-n)\,,\label{eq:aVM:G}\\
& \nabla \cdot B = 0\,, \label{eq:aVM:T}\\
& E = \tilde{E} - \nabla p \,. \label{eq:aVM:Corr}
 \end{empheq}  
\end{subequations}
The right hand side of Eq.~\eqref{eq:aVM:G} can be interpreted as the consistency default in Gauss's law. The corrector $p$ vanishes, subject to the boundary conditions, as soon as the electric field advanced thanks to the Amp\`ere equation verifies the Maxwell-Gauss law. 

The difficulty in handling the quasi-neutral limit is thus twofold. In addition to the degeneracy of the Amp\`ere equation \eqref{eq:aVM:A}, a means of computing the corrector needs to be worked out for the limit regime, the Maxwell-Gauss equation \eqref{eq:aVM:G} also degenerating in the quasi-neutral limit. This last difficulty is similar to the one posed by the computation of the electric potential in the electrostatic framework. The investigation of the augmented Vlasov-Maxwell system is thus a good means of offering a unified presentation of both regimes.

\subsubsection{Reformulation of the augmented Vlasov-Maxwell system}\label{sec:ref:Maxwell}

The objective here is to restore a means of computing the electric field in the quasi-neutral regime. \Fabrice{As mentioned above, in the Maxwell regime, the electric field is computed thanks to the displacement current. This term vanishing from the equation in the limit $\lambda^2\to0$, Ohm's law needs to be considered in order to express how the electric field relates to the current of particles. This finally restores means of computing the electric field in Amp\`ere's law}. Letting $\lambda^2 \to 0$ in \eqref{eq:aVM:A} and taking the formal time derivative of this equation together with the curl of Faraday's law \eqref{eq:aVM:F} yields
\begin{equation}\label{eq:Amp:degenerated}
  \nabla \times \nabla \times E = \frac{\partial J}{\partial t} \,.
\end{equation}
In this equation a link between the electric field and the electric sources is restored. However, it does not allow for the computation of the entire electric field. Indeed, the solution of this equation can be augmented by any gradient without changing the equality: the electrostatic component of the field cannot be uniquely determined from \eqref{eq:Amp:degenerated}. This is corrected thanks to the expression of the current of particles which translates the response of the particles to the electric field, with
\begin{equation*}
  \frac{\partial J}{\partial t} = \nabla \cdot \mathbb{S} + n \tilde E - J \times B \,.
\end{equation*}
Inserting this definition into \eqref{eq:Amp:degenerated}, the quasi-neutral equation providing the entire electric field in the quasi-neutral regime can be precised, with
\begin{equation*}
   \nabla \times \nabla \times \tilde E + n \tilde E = J\times B - \nabla \cdot \mathbb{S} \,.
\end{equation*}

A similar reformulation can be performed for the correction potential $p$. Indeed, Eq.~\eqref{eq:aVM:G} degenerates into the quasi-neutrality relation $1-n = 0$. This constraint is operated together with the moments of the Vlasov equation in order to derive the equation verified by the corrector. Following the spirit of the Boris procedure, a correction of the electric field is introduced in order for the continuity equation to be verified. 
Taking the double time derivative of this equation together with the moments of the Vlasov equation, in which the electric field is corrected, the following equation is derived
\begin{equation}
  \frac{\partial^2 n}{\partial t^2} = \nabla \cdot \frac{\partial J}{\partial t} = \nabla^2: \mathbb{S} + \nabla\cdot \big( n (\tilde{E} - \nabla p)\big) - \nabla \cdot ( J \times B )\,,
\end{equation}
where $ \nabla^2: \mathbb{S} := \nabla \cdot ( \nabla \cdot \mathbb{S})$. 
 This finally provides the equation verified by the corrector in the limit $\lambda^2\to0$, so that it is possible to state the quasi-neutral Vlasov-Maxwell system
 \begin{subequations}\label{eq:aVM0} 
 \begin{empheq}[left=(aVM)^0 \empheqlbrace]{align}
&\frac{\partial f}{\partial t} + v \cdot \nabla_x f - ({E}+  v \times B)\cdot \nabla_v f =0\label{eq:aVM0:V}\\
& \nabla \times \nabla \times \tilde E + n \tilde E = J\times B - \nabla \cdot \mathbb{S} \,,\label{eq:aVM0:A}  \\
&  \frac{\partial J}{\partial t} + \nabla \times \tilde{E} = 0,\label{eq:aVM0:F}\\
\begin{split}
&- \nabla \cdot (n \nabla p) = \frac{\partial ^2 n}{\partial t^2} - \nabla^2:\mathbb{S} - \nabla \cdot (n \tilde{E}) + \nabla \cdot ( J \times B) \,, \label{eq:aVM0:G}
\end{split} \\
& \nabla \cdot B = 0\,, \label{eq:aVM0:T}\\
& E = \tilde{E} - \nabla p \,. \label{eq:aVM0:Corr}
 \end{empheq}  
\end{subequations}
In the quasi-neutral limit, the electric field can be interpreted as the Lagrange multiplier of the equilibrium $\nabla \times B = J$, the corrector potential as the Lagrange multiplier of the constraint $\nabla \cdot J = 0$, or, more precisely to the time derivative of these identities.
The equation \eqref{eq:aVM0:V} outlines the singular nature of the quasi-neutral limit: the electric field verifies an hyperbolic equation in the Maxwell regime defined in section \ref{sec:QN:intro} while it is computed thanks to an elliptic equation in the quasi-neutral limit. On top of that, these two equations relate different physical phenomena, the propagation of an electromagnetic wave at the speed of light on the one hand, the response of the charged particles to the electric field on the other hand.
The quasi-neutral regime investigated with this limit model is close to a kinetic description of the so-called Electron MHD \cite{gordeev_electron_1994} \Fabrice{and the quasi-neutral model identified in \cite{tronci_neutral_2015}}. In this system the scale of interest is that of the electron, rather than the ion dynamics in the classical MHD models, with a finite electron inertia. Moreover the model defined by \eqref{eq:aVM0} remains a fully kinetic description for the plasma.
\medskip

The aim of the reformulation, leading to an asymptotic preserving method, is to bring these two regimes into a single set of equations with a smooth transition from one to the other one according to the values of $\lambda$. With this aim, a derivation similar to that of the limit problem \eqref{eq:aVM0} is performed but keeping $\lambda>0$. This yields the reformulated Vlasov-Maxwell system
 \begin{subequations}\label{eq:RaVM} 
 \begin{empheq}[left=(RaVM)^\lambda \empheqlbrace]{align}
& \frac{\partial J}{\partial t} + v \cdot \nabla_x f -  ({E}+ v \times B)\cdot \nabla_v f =0\label{eq:RaVM:V}\\
&\lambda^2  \frac{\partial^2 \tilde{E}}{\partial t^2} +  \nabla \times \nabla \times \tilde E + n \tilde E = J\times B - \nabla \cdot \mathbb{S} \,,\label{eq:RaVM:A}  \\
&  \frac{\partial J}{\partial t} + \nabla \times \tilde{E} = 0,\label{eq:RaVM:F}\\
\begin{split}
&- \lambda^2 \frac{\partial^2}{\partial t^2} \Delta p - \nabla \cdot (n \nabla p) = \\
&\hspace*{2cm} \frac{\partial ^2 n}{\partial t^2} - \nabla^2:\mathbb{S} - \nabla \cdot (n \tilde{E}) + \nabla \cdot ( J \times B) \,, \label{eq:RaVM:G}
\end{split} \\
& \nabla \cdot B = 0\,, \label{eq:RaVM:T}\\
& E = \tilde{E} - \nabla p \,. \label{eq:RaVM:Corr}
 \end{empheq}  
\end{subequations}

\begin{remark}
    \begin{enumerate}[a)]
    \item The reformulated Amp\`ere equation \eqref{eq:RaVM:A} is well posed (provided adequate boundary conditions) for all values of $\lambda^2$. Indeed in the limit $\lambda \to 0$ the plasma density is large and the operator $\nabla \times \nabla \times E + n E$ is elliptic. Conversely, when $n \to 0$ the scaled Debye length is large and the equation remains well posed. These remarks also apply to the reformulated Gauss law \eqref{eq:RaVM:G} providing the corrector.
    \item The quasi-neutral Vlasov-Maxwell system \eqref{eq:aVM0} is recovered from the reformulated system when $\lambda \to 0$. The quasi-neutral limit is a regular perturbation of the reformulated system \eqref{eq:RaVM}.
    \item The right hand side of the equation~\eqref{eq:RaVM:G} can be interpreted as the default of consistency with the continuity equation a common feature with the Boris correction \cite{boris_proceedings_1970}. In this respect, this equation can be regarded as a generalization of the Boris correction.
    \end{enumerate}
\end{remark}

\subsection{Overview of the numerical methods}\label{sec:QN:num}

The purpose here is to use the concepts introduced in the precedent section for the continuous system and to transpose them to the discrete equations. Generally, the time discretization is a key point in the derivation of an AP numerical method. Due to the singular nature of the quasi-neutral limit several quantities must be computed thanks to an implicit time discretization in order to secure the consistency with both the Maxwell and the quasi-neutral regimes and to provide a means of computing the electric field in every regime. 

 The level of implicitness is controlled by three parameters  
 $(a,b,c)$, the value of each one being equal to either 1 or 0 \cite{degond_numerical_2012}. 
\begin{subequations}\label{eq:system:semi:discret:temps}
\begin{eqnarray}
& & \hspace{-1cm} \frac{1}{\Delta t} (n^{ m+1} - n^{ m}) - \nabla \cdot J^{ m+a} = 0, \label{DS1F_n} \\
& & \hspace{-1cm}  \frac{1}{\Delta t} ( J^{ m+1} - J^{ m}) - \nabla \cdot \mathbb{S}^m  =   n^{ m+1-a} E^{ m+1} - J^m \times B^{ m}, \label{DS1F_u} \\
& & \hspace{-1cm} \frac{1}{\Delta t} (B^{ m+1} - B^{ m}) + \nabla \times E^{ m+b} = 0, \label{DS1F_B} \\
& & \hspace{-1cm} \lambda^2 \frac{1}{\Delta t} (\tilde{E}^{ m+1} - E^{ m}) - \nabla \times B^{ m+c} =  - J^{ m+a}  , \label{DS1F_E} \\
& & \hspace{-1cm} \lambda^2 \nabla \cdot E^{ m+1} = (1 - n^{ m+1})\,, \label{DS1F_divE}\\  
& & \hspace{-1cm} E^{ m+1} =  \tilde{E}^{ m+1} - \nabla p \,.\label{DS1F_p}
\end{eqnarray}
supplemented with  $ \nabla \cdot B^{ m+1} = 0$.
\end{subequations}


At this stage, different remarks can be stated:
\begin{enumerate}[a)]
\item The quasi-neutral regime is recovered for vanishing $\lambda$ which stands for both the scaled Debye length and the ratio of the typical velocity to the speed of light. The stability with respect to $\lambda$ requires therefore an implicit discretization of the (homogeneous) Maxwell equations, yielding to $b=c=1$. This is related to the assumption that the typical velocity is small compared to the speed of light ($\alpha\to0$)
\item The consistency property with respect to the quasi-neutral regime requires an implicit particle current $J$ in Amp\`ere's law \eqref{DS1F_n} with and implicit electric field in the definition of $J$. Accordingly, an implicit electric field must be used in the Lorentz force defining the source of the momentum equation Eq.~\eqref{DS1F_u}. These requirements are met for $a=1$. Note that the scaling assumptions imply that the dimensionless Debye length also represents the scaled plasma period. Therefore, the uniform stability property with respect to $\lambda$ brings the stability of the method for time steps lager than the plasma period.
\item The density occurring in the Lorentz force is made explicit when the mass flux is implicit in order to uncouple the resolution of the Eqs.~\eqref{DS1F_n} and \eqref{DS1F_u}.
\item The consistency with the Maxwell-Gauss equation at the discrete level, requires the same level of implicitness for the mass flux in Eq.~\eqref{DS1F_n} and the current $J$ in Amp\`ere's law \eqref{DS1F_E}. This point will be detailed further in the sequel.
\end{enumerate}
The linear stability proposed in \cite{degond_numerical_2012} demonstrates that the AP property cannot be achieved with an implicitness level weaker than $(a,b,c)=(1,1,1)$. \Fabrice{This choice defines a consistent discretization of the reformulated system \eqref{eq:RaVM}. Indeed, Eqs.~\eqref{DS1F_B}, \eqref{DS1F_E} and \eqref{DS1F_u} in which the correction is omitted yield
\begin{multline}\label{eq:proof:consistancy}
  {\frac{\lambda^2}{\Delta t^2} \left( \tilde{E}^{m+1} - {E}^{m} \right) = \frac{1}{\Delta t}\Big(\nabla \times B^m - J^m \Big)}\\ - {\nabla \times \nabla \times {\tilde{E}^{m+1}} -{n}^m\tilde{E}^{m+1} - \nabla\cdot S^m + J^m \times B^m \,.}
\end{multline}
Owing that Amp\`ere's law is initially verified: $
  \nabla \times B^m - J^m  \approx \frac{\lambda^2}{\Delta t} \left(E^m - E^{m-1} \right)$, the following identity holds
\begin{multline*}
  \frac{\lambda^2}{\Delta t^2} \left(\tilde{E}^{m+1} - 2 {E}^{m} + E^{m-1}\right) + \nabla \times \nabla \times\tilde{E}^{m+1} + \\
  {n}^m\tilde{E}^{m+1} + \nabla\cdot S^m - J^m \times B^m \approx 0 \,,
\end{multline*}
which defines a time semi discretization of the reformulated Amp\`ere equation \eqref{eq:RaVM:A}. A similar result can be obtained for the Gauss law, with Eqs.~\eqref{DS1F_n}, \eqref{DS1F_u}, \eqref{DS1F_divE}, \eqref{DS1F_p} and \eqref{eq:proof:consistancy} providing
\begin{multline*}
  -\nabla \cdot \Big(\big(\frac{\lambda^2}{\Delta t ^2} +  n^{m}\big) \nabla p\Big) =  \frac{1}{\Delta t^2} \Big( 1 -\tilde{n}^{m+1} - \lambda^2 \nabla \cdot E^{m} \Big) +\frac{1}{\Delta t}\nabla \cdot  J^m \\ + \nabla^2 : S^m - \nabla\cdot ( J^m \times B^m ) + \nabla \cdot  ( {n}^m\tilde{E}^{m+1} ) \,,
\end{multline*}
where $\tilde{n}^{m+1} = {n}^{m+1}+\Delta t^2 \nabla \cdot ( n^m \nabla p)$. 
Assuming that Gauss law and the continuity equation are satisfied at the previous time step $  \lambda^2 \nabla \cdot E^m \approx 1 -  {n}^{m}$ and $ \Delta t\nabla \cdot  J^m \approx  {n}^{m} - {n}^{m-1}$
the following identity holds
\begin{multline*}
  -\nabla \cdot \Big(\big(\frac{\lambda^2}{\Delta t ^2} +{n}^{m}\big) \nabla p\Big) \approx \frac{1}{\Delta t^2} \Big( -\tilde{n}^{m+1} +2 {n}^{m} - {n}^{m-1} \Big) \\+\nabla^2 : S^m - \nabla\cdot ( J^m \times B^m ) + \nabla \cdot  ( n^m\tilde{E}^{m+1}) \,.
\end{multline*}
This defines a time discretization of the reformulated Gauss law \eqref{eq:RaVM:A} provided that the correction at time level $m$ and $m-1$ vanishes.}

\Fabrice{\begin{remark}\label{rem:form:AP}
The time discretization is operated in a way to avoid the time differentiation of the equations and thus provide the consistency with Amp\`ere's equation and the time derivative of Gauss's law rather than with their time derivatives as suggested by the continuous reformulated system \eqref{eq:RaVM}. This remark is clearly illustrated by comparing the reformulated Ampere equation \eqref{eq:RaVM:A}  with its discrete counter part \eqref{eq:proof:consistancy}. The former incorporates a double time derivative of the electric field, the latter, exploiting the time discretization, avoids the time differentiation of Amp\`ere's law. 
This explains the difference between the PIC-AP1 and PIC-AP2 methods proposed in \cite{degond_asymptotically_2006,degond_asymptotic-preserving_2010} for the Vlasov-Poisson system. The first one is derived as a discretization of the reformulated continuous system and is thus consistent with the double time derivative of Gauss's law. This is in line with Eq.~\eqref{eq:RaVM:G}. The second one is derived thanks to the discrete set of equations and, working the time discretization, implements a parabolic equation rather than a wave like equation. This gains an advantage since only one initial condition is necessary.
\end{remark}}

\medskip

Different space discretization have been considered. For kinetic description, either Particle-In-Cell \cite{degond_asymptotically_2006,degond_asymptotic-preserving_2010,DegDelDoy} or semi-Lagrangian \cite{BCDS09} discretizations have been implemented, while, for fluid descriptions finite volume (on Cartesian meshes) are used \cite{CDV07,degond_numerical_2012}. In this last series of works dedicated to the Euler-Maxwell system, an exact consistency with the Maxwell-Gauss equation can be obtained. To this end, the numerical flux associated to the mass flux must be used to construct the current of particles used in the Amp\`ere equation. This property is sketched in the next lines in a simplified one-dimensional framework, with $B_x=0$. Denoting  $\mathcal{F}^{m+1}_{k+1/2}$ the numerical flux associated to the mass flux evaluated at the center of the cell $k$, $n_{k}^{m}$ and  $E_x^m|_{k+1/2}$ being the density and the electric field at time $t^{m}= m \,\Delta t$, with $\Delta t$ and $\Delta x$  the time and space mesh intervals, a discretization of the system \eqref{DS1F_n} and \eqref{DS1F_E} is written
\begin{align}
  n_k^{m+1} = n_{k}^{m} + \frac{\Delta t}{\Delta x} \left(\mathcal{F}^{m+1}_{k+1/2} - \mathcal{F}^{m+1}_{k-1/2} \right)\,, \label{1F_n_fd}\\
  \lambda^2 \frac{1}{\Delta t} (E_x^{m+1}|_{k+1/2} - E_x^m|_{k+1/2}) = \mathcal{F}^{m+1}_{k+1/2} \,.\label{1F_amperex_fd}
\end{align}
Eq.~\eqref{1F_amperex_fd} evaluated at the cell interfaces $x_{k+1/2}$ and $x_{k-1/2}$ together with (\ref{1F_n_fd}) yields to
\begin{equation*}
 \lambda^2  \frac{1}{\Delta x} (E_x|_{k+1/2}^{m+1} - E_x|_{k-1/2}^{m+1}) + n|_{k}^{m+1} = \lambda^2  \frac{1}{\Delta x} (E_x|_{k+1/2}^{m} - E_x|_{k-1/2}^{m}) + n|_{k}^{m}.  \label{1F_gauss_fd}
\end{equation*}
This expression defines a discretization of the equation $\frac{\partial }{\partial t} \left( \lambda^2 \partial E_x/\partial x  \right) =  - \frac{\partial n}{\partial t}  $.

\medskip

A similar property cannot be obtained with standard PIC methods, the macroscopic quantities projected on the grid are indeed inconsistent with the Gauss law. Therefore the correction is mandatory. The time discretization of the equation providing this quantity is straightforwardly obtained from the system \eqref{eq:system:semi:discret:temps}.

The scaling relations defining the quasi-neutral regime mean that beside the dimensionless Debye length and the ratio of the typical velocity to the speed of light, the asymptotic parameter $\lambda$ carries the scaled plasma period $\tau_p$ as well. Indeed, the following identity  $\lambda^2 = \lambda_D^2/\bar x^2 = {M^2}/{(\bar t \omega_p)^2}$, $\omega_p=1/\tau_p$ being the plasma frequency, together with the assumption $M=1$ proves the above assertion. 
The plasma period usually defines one of the smallest time scales involved in plasma modelling. To perform simulations on large scales, implicit methods have received a lot of attention, specifically in the framework of PIC discretizations for kinetic plasma models with the direct implicit methods \cite{LCF83,CLF82,CLHP89,HeLa87} or the moment implicit methods \cite{Mas81,BrFo82,WBF86,Mas87,RLB02}.
The uniform stability with respect to $\lambda$ ensures that AP-methods remain stable for discretization that do not resolve the plasma period.  Therefore, AP methods share some analogies with implicit or semi implicit methods. We refer to \cite[section~4]{DegDelDoy} for a more thorough discussion.


\section{Drift limit for fluid descriptions of hot plasmas under large magnetic fields}
\label{sec:drift}

\subsection{Introduction}

This section is devoted to the design of fluid models and numerical methods for thermal plasmas evolving under a strong magnetic field. The targeted applications are tokamak plasmas and magnetically confined fusion \cite{chen_indispensable_2011,miyamoto_plasma_2005,stacey_fusion_2005,freidberg_plasma_2007}. Its principle consists in heating a plasma to hundreds of thousands of degrees in order for the thermal agitation to overcome the Coulomb repulsion. Indeed, for the nuclear reaction to occur, the distance between the nuclei has to be lower than $10^{-15}$ m. Simultaneously, the plasma expansion is prevented by confining the particles thanks to an intense magnetic field. 

This section presents asymptotic preserving methods developed for simulating these hot plasmas evolving in the presence of strong magnetic fields. The magnetization of the plasma induces a severe anisotropy, with different parallel and perpendicular (with respect to the magnetic field) dynamics. The dynamic of interest is the transverse one, which is, compared to the parallel dynamic, a low frequency regime driven by the drift waves. Along the magnetic field lines, the acoustic waves are very fast, balancing almost instantly the electric field in order to enforce a zero force regime. The first purpose of the AP methods described in this section is to follow the slow perpendicular dynamic while accounting for the parallel force balance as well.
The second aim is to explore the effectiveness of numerical methods that do not use coordinate systems adapted to the magnetic field geometry by contrast to standard approaches (see for instance \cite{beer_fieldaligned_1995,dimits_fluid_1993,dudson_bout++:_2009,hammett_developments_1993,ottaviani_alternative_2011,ricci_simulation_2012,scott_shifted_2001,scott_free-energy_2005,balescu_transport_1988,hazeltine_plasma_2003}). Indeed standard approaches have some difficulties in specific areas such as the so-called ``O point'' or ``X point'' \cite{hariri_flux-coordinate_2013,f_hariri_flux-coordinate_2014,shanahan_x-point_2014} where the coordinate system is singular. Moreover, the goal pursued here is to easily account for changes in the magnetic field topology, with for instance the creation of magnetic islands \cite{biskamp_magnetic_2005,bellan_fundamentals_2008,hazeltine_plasma_2003,hill_effect_2015,deluzet_numerical_2015,wesson_finite_1966,rutherford_nonlinear_1973,winske_hybrid_1994}. To this end, the numerical methods developed within this section are free from any assumption relating the magnetic field to the mesh or the coordinate system.

The collisions within tokamaks are weak, therefore the reliability of fluid descriptions is questioned \cite{grandgirard_drift-kinetic_2006} especially in the case of devices such as ITER and DEMO (see the main characteristics in table~\ref{TAB:tokamak:ITER}), with investigations \cite{dimits_comparisons_2000} demonstrating that fluid models underestimate instability thresholds and overestimate turbulent fluxes. However, numerous fluid descriptions are implemented for tokamak plasmas \cite{lutjens_xtor_2008,huysmans_non-linear_2009}, in particular for the study of the plasma edge physics (see for instance \cite{dudson_bout++:_2009,tamain_tokam-3d:_2010,bufferand_study_2014}). Additionally, kinetic corrections may be formulated for fluid models in order to correct some of their flaws \cite{grandgirard_drift-kinetic_2006}. The main argument, however, seems to be the huge cost needed for the numerical resolution of kinetic models which is challenging in term of computational time as well as memory usage. This is the ``curse of dimensionality'': even in the gyro-kinetic approximation \cite{littlejohn_variational_1983,garbet_gyrokinetic_2010,bufferand_study_2014} kinetic models consider a distribution function in a five dimensional phase-space. Though their accuracy may be improved, fluid models provide access to a rich physics which helps understanding the various regimes that prevail within tokamaks and allows for the study of instability mechanisms \cite{mikhailovskii_instabilities_1998, stacey_fusion_2005, biskamp_nonlinear_1997}.
\begin{table}[!ht]
  \centering
  \caption{Characteristics of the ITER and DEMO tokamaks \cite{maisonnier_power_2007,freidberg_plasma_2007}.}\label{TAB:tokamak:ITER}
  \begin{tabular}{|l|c|c|}\hline
    & ITER & DEMO \\ \hline \hline
    External radius (m) & 6.2 & 6.1-9.55 \\ \hline 
    Inner radius (m) & \multicolumn{2}{c|}{2} \\ \hline \hline
    Magnetic field intensity (T) & 5.3 & 7 \\ \hline \hline
    Plasma density (m$^{-3}$) & $10^{20}$ & 1.5 $10^{20}$\\\hline 
    Plasma Temperature (eV) & \multicolumn{2}{c|}{$10^4$} \\\hline \hline
    Discharge duration (s) & 400 & 1000 \\\hline
  \end{tabular}
\end{table}

 The purpose of this section is thus to detail the Asymptotic-Preserving methodology in this framework. The plasma is described by two sets of fluid equations, the aim being to investigate the drift limit also referred to as gyro-fluid regime \cite{degond_asymptotic-preserving_2011}. The drift asymptotic shares some analogies with the low-Mach regime and the difficulty investigated here consists in the transition between this low Mach flow which prevails in the plasma core, to regimes with an increased mean velocity and a lower temperature. This dynamical transition is observed in the Scrape-Off Layer (SOL), precisely in the magnetic pre-sheath and the sheath where the plasma temperature drops and the ions are accelerated by the electrostatic field \cite{post_physics_1986,stangeby_plasma_2000}, the flow being potentially supersonic \cite{chodura_plasma_1986}.

Two difficulties need to be overcome in order to efficiently address the drift regime in the context of tokamaks. The first one relates to the vanishing of the inertia which defines a zero force regime. Along the magnetic field lines, the parallel momentum equation is degenerate which prevents from computing the parallel momentum explicitly. A first approach is proposed in order to overcome this difficulty. The aligned momentum component is considered as the Lagrange multiplier of the parallel force balance. An alternative is constructed on ideas borrowed from the low Mach regime \cite{degond_all_2011,haack_all-speed_2012} with a computation of the pressure securing the zero force regime along the magnetic field lines. The singularity of the problem is thus overcome by the resolution of an anisotropic diffusion problem along the aligned direction.
The second difficulty addressed here is related to the periodicity of the torus containing the plasma. The anisotropic diffusion equations derived to transform the drift limit into a regular limit are supplemented with periodic boundary conditions at each end of the magnetic field lines. This defines an ill posed problem  in the limit regime. This is a characteristic problem of tokamaks, with highly magnetized plasmas evolving in a periodic geometry. For this class of anisotropic problems, Asymptotic-Preserving techniques have been developed. Their presentation is postponed to Section~\ref{sec:aniso}.

This section is organized as follows. The drift regime for the Euler-Lorentz system is stated, with the definition of the scaling relations. The magnetic field is assumed to be a given data that does not depend on time. Two Asymptotic-Preserving reformulations are thus proposed with, finally, the detailed computation of the self-consistent electric field.

\def\q{\mathfrak{q}}
\def\Id{\mathbb{I}\text{d}}

\subsection{The Euler-Lorentz model in the drift regime}\label{sec:EL:drift}

The plasma is described by a bi-fluid model, consisting of a system of compressible Euler equations for the ions and the electrons. The density, momentum and energy associated to electrons and the ions are denoted 
$(n_\alpha,q_\alpha,W_\alpha)$ ($\alpha=e$ for the electrons and $i$ for the ions), the charge of the particle being denoted $\q_\alpha$. These quantities verify the Euler-Lorentz system
\begin{subequations}\label{eq:EL:Alpha}
  \begin{align}
    & \frac{\partial n_\alpha}{\partial t} + \nabla \cdot q_\alpha = 0 \,, \\
    & m_\alpha \left(\frac{\partial q_\alpha}{\partial t} + \nabla \cdot \left( q_\alpha \otimes \frac{q_\alpha}{n_\alpha} \right) \right) + \nabla p_\alpha = \q_\alpha \left(n_\alpha E + q_\alpha \times B \right) \,, \\
    & \frac{\partial W_\alpha}{\partial t} + \nabla \cdot \Big( (W_\alpha+p_\alpha)\frac{q_\alpha}{n_\alpha} \Big) = \q_\alpha E\cdot q_\alpha \,, 
  \end{align}
where $k_B$ is Boltzmann constant, $m_\alpha$ the particle mass, $T_\alpha$ and $p_\alpha$ the temperature and the pressure of the gas, with the following expression of the energy
\begin{equation}
  W_\alpha = \frac{1}{2} m_\alpha  \frac{|q_\alpha|^2}{n_\alpha} + \frac{3}{2} p_\alpha \,, \qquad p_\alpha = n_\alpha k_B T_\alpha \,.
\end{equation}
\end{subequations}
The given magnetic field is assumed static, i.e. satisfying
\begin{align*}
  &\frac{\partial B}{\partial t} = 0 \,, \qquad \nabla \cdot B = 0 \,.
\end{align*}  

The electronic and ionic conservation equations are coupled to the electro-static field defined as 
  \begin{equation}\label{eq:EL:Poisson}
  - \Delta \phi = \frac{\q}{\varepsilon_0} (n_i - n_e) \,, \qquad \text{with } E = - \nabla \phi \,,
  \end{equation}
with $\varepsilon_0$ the vacuum permeability and $\q$ the elementary charge.

In order to identify the drift regime, the system is rewritten using dimensionless variables. Denoting $\bar n$ the typical value for the plasma density, the identity $n = \bar n n'$ holds for the physical ($n$) and dimensionless ($n'$) quantities. The space and time scales of the observed phenomenon are $\bar x$ and $\bar t$, defining $\bar u = \bar x / \bar t$ the typical velocity.  Different dimensionless parameters will be used, with $\varepsilon$ the electronic to the ionic mass ratio, $\lambda$ the Debye length scaled by $\bar x$, $\bar t \omega_c$ the number of cyclotron period during the typical time, $M$ the ionic Mach number, $c_s$ denoting the ionic speed of sound, and $\bar u $ the typical mean plasma velocity, 
\begin{align*}
  &\varepsilon = \frac{m_e}{m_i}\,, \\
    &\lambda = \frac{\lambda_D}{\bar x}\,, \qquad \lambda_D^2 = \frac{\varepsilon_0 k_B \bar T}{\q^2 \bar n} = \frac{\varepsilon_0 \bar p}{\q^2 \bar n^2} \,, \\
  &\bar t \omega_c = \bar t \, \frac{\q B}{m_i} \,,\\
   & M^2 = \frac{\bar u^2}{c_s^2}\,, \qquad c_s^2 = \frac{\bar p}{m_i \bar n} \,, \qquad \bar u = \frac{\bar q}{\bar n} \,.
\end{align*}

The number of free parameters is reduced thanks to scaling relations defining the regime in which the system is observed. The space and time scales introduced define a typical velocity which is assumed to be comparable to the mean flow velocity, the electric and magnetic field verify a scaling relation characterizing a quasi-neutral regime, similar to the frozen field assumption described in section \ref{sec:QN:intro}. Finally, the Mach number and the dimensionless cyclotron period are assumed to define to small scales unified in a single parameter denoted $\tau$.

 These assumptions can be summarized as follows 
\begin{equation}\label{eq:drift:scaling:relations}
    \bar u = \frac{\bar x}{\bar t}\,, \qquad     \bar E = \bar u \bar B \,, \qquad    \frac{1}{M^2} = \bar t \omega_c = \frac{1}{\tau}\,, 
  \end{equation}

The Euler-Lorentz-Poisson system (\ref{eq:EL:Alpha}--\ref{eq:EL:Poisson}) can thus be written using  three dimensionless parameters $\varepsilon$, $\lambda$ and $\tau$, with, for the ions, 
\begin{subequations}\label{eq:EL:ion:Adim}
  \begin{align}
    & \frac{\partial n}{\partial t} + \nabla \cdot q = 0 \,, \label{eq:EL:ion:Adim:a}\\[3mm]
    & \frac{\partial q}{\partial t} + \nabla \cdot \left( q \otimes \frac{q}{n} \right) 
+ \frac{1}{\tau} \nabla p =
    \frac{1}{\tau}  \left( n E + q \times B \right) \,,  \label{eq:EL:ion:Adim:b}    \\ 
    & \frac{\partial W}{\partial t} +  \nabla \cdot \big( (W+p) \frac{q}{n} \big)\big) =  E\cdot q \,, \label{eq:EL:ion:Adim:c} \\
&  W = \tau  \frac{1}{2}\frac{q^2}{n} + \frac{3 p}{2}\,, \qquad p =  n T \,.
\end{align}                
\end{subequations}
A similar system is written for the electrons
\begin{subequations}\label{eq:EL:elec:Adim}
  \begin{align}
    & \frac{\partial n_e}{\partial t} + \nabla \cdot q_e = 0 \,, \label{eq:EL:elec:Adim:a}\\[3mm]
    & \varepsilon\left( \frac{\partial q_e}{\partial t} + \nabla \cdot \left( q_e \otimes \frac{q_e}{n_e} \right)\right) + \frac{1}{\tau} \nabla p_e =
   - \frac{1}{\tau} \left( n_e E + q_e \times B \right) \,,      \\ 
    & \frac{\partial W_e}{\partial t} +  \nabla \cdot \big( (W_e+p_e)\frac{q_e}{n_e} \big)\big) = -E\cdot q_e \,,
  \end{align}
  \begin{align}
    & W_e = \varepsilon \tau \frac{1}{2} \frac{q_e^2}{n_e} + \frac{3 p_e}{2}\,, \qquad p_e =  n_e T_e \,.
  \end{align}
\end{subequations}
The electric field is provided by
\begin{equation}\label{eq:poisson:adim}
  -  \lambda^2\Delta \phi = n - n_e \,, \qquad   E = - \nabla \phi \,.
\end{equation}

Using the scales reported in the table \ref{TAB:tokamak:ITER} the dimensionless Debye length is evaluated to be as small as $10^{-5}$ which, by the equation \eqref{eq:poisson:adim}, brings the quasi-neutrality assumption. This is consistent with the scaling considered for the electro-magnetic field as discussed above. Another consequence of the quasi-neutral regime is the degeneracy of the Poisson equation \eqref{eq:poisson:adim} that cannot be used to compute the electric potential in this regime. Two alternatives have been proposed, and detailed in the sequel (see section~\ref{sec:calcul:E}). Finally, the Mach number and the scaled cyclotron period are estimated equal to $10^{-8}$ and $10^{-6}$ respectively. These quantities define thus a small scale characterizing the evolution of hot plasmas under intense magnetic fields at the origin of severe difficulties for efficient numerical simulations. To highlight these difficulties, some notations need to be introduced.
 \begin{itshape}
\paragraph{Notations} $B$ denoting the magnetic field, $b$ the unit vector pointing in the direction of $B$, with $ b := {B}/{|B|}$,  
for all scalar $p$ and vector $q$, we define
\begin{alignat}{4}
  &q_\parallel := (q\cdot b) \,,&& \qquad q_\perp :=  q - b q_\parallel = (\Id - b\otimes b) q = b \times q \times b\,,\notag\\[3mm]
  &\nabla_\parallel p :=  b\cdot\nabla p  \,,&& \qquad \nabla_\perp p := (\Id - b\otimes b) \nabla p = \nabla p - (\nabla p \cdot b) b  \,,\label{eq:notations:B}\\[3mm]
  &\nabla_\parallel \cdot q_\parallel  := \nabla \cdot (q_\parallel \, b ) \,,&& \qquad \nabla_\perp \cdot q_\perp := \nabla \cdot (q_\perp) \,.\notag
\end{alignat}  
 \end{itshape}
Inserting formally $\tau =0$ into \eqref{eq:EL:ion:Adim} yields to
\begin{subequations}\label{eq:EL:ion:deriv:zero}
  \begin{align}
    & \frac{\partial n}{\partial t} + \nabla \cdot q = 0 \label{eq:EL:ion:deriv:zero:a} \,, \\[3mm]
    & \nabla p = nE + q\times B \label{eq:EL:ion:deriv:zero:b}\,,\\ 
    & \frac{\partial W}{\partial t} +  \nabla \cdot \big( (W+p)\frac{q}{n} \big)\big) =  E\cdot q \,, \qquad W = \nicefrac{3}{2} \ p\,.\label{eq:EL:ion:deriv:zero:d}
  \end{align}
\end{subequations}

Eq.~\eqref{eq:EL:ion:deriv:zero:b} translates a ``zero force regime'', that does not allow for the computation of the aligned momentum. Indeed, projecting onto the parallel and perpendicular direction, the following balances occur
\begin{subequations}\label{eq:zero:force}
\begin{align}
   &q_\perp = \frac{b}{|B|} \times \left( n E + \nabla p\right)\,,\label{eq:zero:force:a}\\      
     &\nabla_\parallel p = n E_\parallel \,,\label{eq:zero:force:b}
\end{align}  
\end{subequations}
In the drift regime, the transverse momentum component $q_\perp$ instantly adjusts to cancel the perpendicular forces. This defines the two classical drift velocities, namely the ``$E \times B$'' drift $(E\times B)/|B|^2$ and the diamagnetic drift $-(\nabla p \times B)/(n|B|^2)$. 

Along the magnetic field lines, the electric field balances the pressure gradient and the system becomes singular for the computation of the parallel momentum. To understand more precisely the parallel dynamic, it is informative to establish the acoustic wave equation.

Using Eq.~\eqref{eq:EL:ion:Adim:c} together with \eqref{eq:EL:ion:Adim:b}, provides an equation for the ionic pressure
\begin{equation*}
  \frac{3}{2} \frac{\partial^2 p}{\partial t^2} +  
  \nabla \cdot \Big( H \frac{\partial q_\parallel}{\partial t} \Big) = 
  - \nabla_\perp \cdot \Big( H \frac{\partial q_\perp}{\partial t} \Big)  - \nabla \cdot \left(\frac{\partial  H}{\partial t} q \right)+ \frac{ \partial}{\partial t} \left(E\cdot q\right) - \frac{\partial^2}{\partial t^2}\Big( \frac{\tau (q) ^2}{2\, n} \Big)
  \,,
\end{equation*}
where $H={W+p}/{n}$ is the system enthalpy. Projecting Eq.~\eqref{eq:EL:ion:Adim:b} onto the magnetic field lines gives
\begin{equation*}
  \frac{\partial}{\partial t} q_\parallel+ \nabla \cdot \left( q \otimes \frac{q}{n} \right)_\parallel =   \frac{1}{\tau}\Big( - \nabla_\parallel p + n E_\parallel \Big) \,, 
\end{equation*}
which leads to
\begin{multline}\label{eq:p:wave}
  \frac{3}{2} \frac{\partial^2 p}{\partial t^2} -  \frac{1}{\tau}
  \nabla \cdot \Big( H (b \otimes b) (\nabla p - n E)\Big) = - \nabla \cdot \left( (b \otimes b) \left( q \otimes \frac{q}{n} \right) \right)\\
 - \nabla \cdot \Big( H \frac{\partial q_\perp}{\partial t} \Big)  - \nabla \cdot \left(\frac{\partial  H}{\partial t} q \right)+ \frac{ \partial}{\partial t} \left(E\cdot q\right) - \frac{\partial^2}{\partial t^2}\Big( \frac{\tau (q) ^2}{2\, n} \Big)
  \,.
\end{multline}
This equations reveals a speed of sound along the magnetic field lines scaling as $1/\sqrt{\tau}$, which demonstrate that in the limit $\tau \to 0$ the pressure waves travel at infinite speed to adjust instantly to the electric force, securing thus the force balance.

Regarding the numerical methods, the difficulty posed by this regime is due to the degeneracy of the equation providing the parallel momentum. The Asymptotic-Preserving methods developed in this framework operate a reformulation of the system in order to regularize this singularity.

\subsection{Momentum based reformulation}\label{sec:EL:ref:Vit}

This approach proposed in \cite{degond_numerical_2013} has been developed for isothermal descriptions with one species of particles (the ions being thus assumed at rest) in \cite{degond_asymptotic_2009,brull_degenerate_2012}. It has been brought to a more elaborated context in \cite{ITGDPO}, for a complete set of equations for the ions, incorporating an energy equation and, an adiabatic response for the electron, the so-called Boltzmann relation (see section~\ref{sec:EL:Boltzmann}). 

This approach makes use of the parallel momentum as a Lagrange multiplier associated to the parallel force balance \eqref{eq:zero:force:b}. The method is sketched for the ions, the equation \eqref{eq:EL:ion:Adim:b} providing 
\begin{equation*}
  \frac{\partial^2}{\partial t^2} q_\parallel - \frac{1}{\tau}\Big( -\frac{\partial  \nabla_\parallel p}{\partial t} + \frac{\partial (n E_\parallel)}{\partial t} \Big) = - \frac{\partial}{\partial t}\left(\nabla \cdot \mathbb{S}_\parallel\right) \,, \qquad   \mathbb{S} = q \otimes \frac{q}{n}   \,, 
\end{equation*}
with, thanks to the energy conservation \eqref{eq:EL:ion:Adim:c}, 
\begin{equation}\label{eq:EL:Vitesse:P}
  \begin{split}
 &\frac{\partial  \nabla_\parallel p}{\partial t} =   - \nabla_\parallel \nabla \cdot \Big( \frac{2}{3}H q_\parallel \Big) + \nabla_\parallel\mathcal{G} \,,     \\
 &  \qquad   \mathcal{G} := \frac{2}{3} \Big(- \nabla\cdot(H q_\perp) + E\cdot q - \frac{\partial}{\partial t}\Big( \frac{\tau (q) ^2}{2\, n} \Big) \Big)\,,
\end{split}
\end{equation}
which finally gives a wave like equation for the aligned momentum:
\begin{equation}\label{eq:EL:Vitess}
  \tau \frac{\partial^2}{\partial t^2} q_\parallel - \Bigg( \nabla_\parallel \nabla \cdot \Big( \frac{2H}{3} q_\parallel \Big) \Bigg)  = - \nabla_\parallel\mathcal{G} +\frac{\partial (n E_\parallel)}{\partial t} - \tau \frac{\partial}{\partial t}\left(\nabla \cdot \mathbb{S}_\parallel\right) \,.
\end{equation}

Note that, in the drift limit, Eq.~\eqref{eq:EL:Vitess} guaranties the zero force regime along the magnetic field lines. Indeed, inserting $\tau=0$ into \eqref{eq:EL:Vitess}, we get
\begin{equation*}\label{eq:EL:Vitess:lim}
   - \Bigg( \nabla_\parallel \nabla \cdot \Big( \frac{2H}{3} q_\parallel \Big) \Bigg)  = - \nabla_\parallel\mathcal{G} +\frac{\partial (n E_\parallel)}{\partial t} \,, 
\end{equation*}
which, owing to \eqref{eq:EL:Vitesse:P} gives
\begin{equation}\label{eq:zero:force:dtb}
   \frac{\partial}{\partial t} \left( \nabla_\parallel p - n E_\parallel \right) = 0.
\end{equation}
\begin{remark} The second order operator involved in the equation \eqref{eq:EL:Vitess:lim} is non standard. It is constructed as the gradient of a divergence. However, being applied to a scalar field, it translates the double derivative in the aligned direction. This equation is therefore well posed provided it is supplemented with adequate boundary conditions. 
  \end{remark}

The first implementation of this reformulation is proposed in \cite{degond_asymptotic_2009} for an isothermal description of the plasma with an external constant electric field. In this context, the combination of the density equation and the momentum equations provides the reformulated equation for the parallel momentum. This equation is supplemented with Dirichlet boundary conditions and discretized in a two dimensional computation domain with a magnetic field aligned with one coordinate.

In \cite{brull_degenerate_2012}, this first achievement is generalized to magnetic field geometries uncorrelated to either the coordinate system or the mesh. The Dirichlet boundary conditions are substituted with Neumann ones, reproducing the difficulty to account for the periodicity of the torus. In this work, the computation of the aligned momentum equation requires the resolution of a diffusion problem with a severe anisotropy, which is ill-posed in the limit $\tau \to 0$. This system is solved thanks to the ``differential characterization'' method described in section \ref{sec:AP:diff}.

\begin{figure}[!ht]
  \centering
  \begin{small}
    \subfigure[Density at t=1.\label{fig:ITG:a}]{%
      \includegraphics[width=0.3\textwidth]{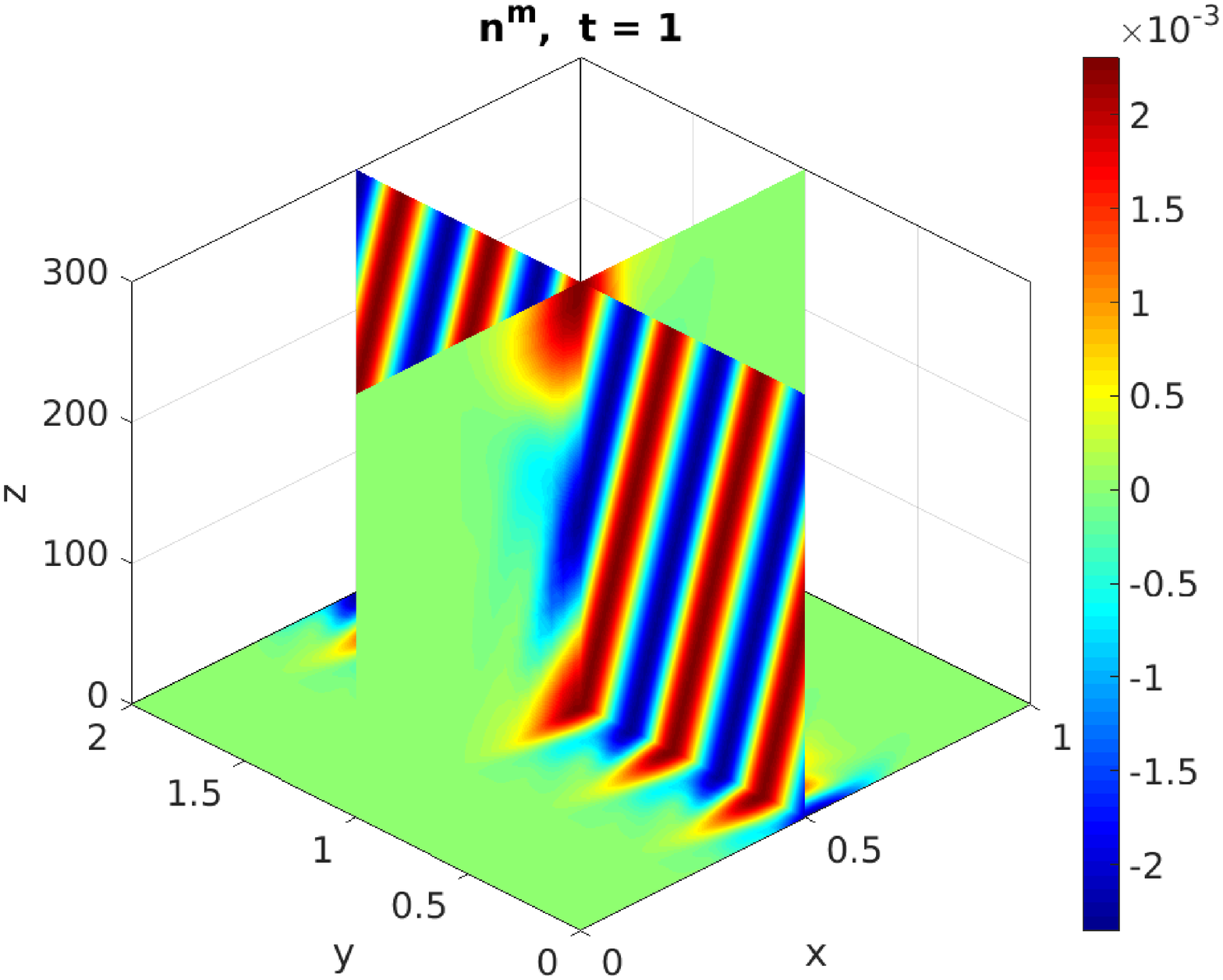}}%
    \hspace*{0.05\textwidth}%
    \subfigure[Density at t=2.\label{fig:ITG:b}]{%
      \includegraphics[width=0.3\textwidth]{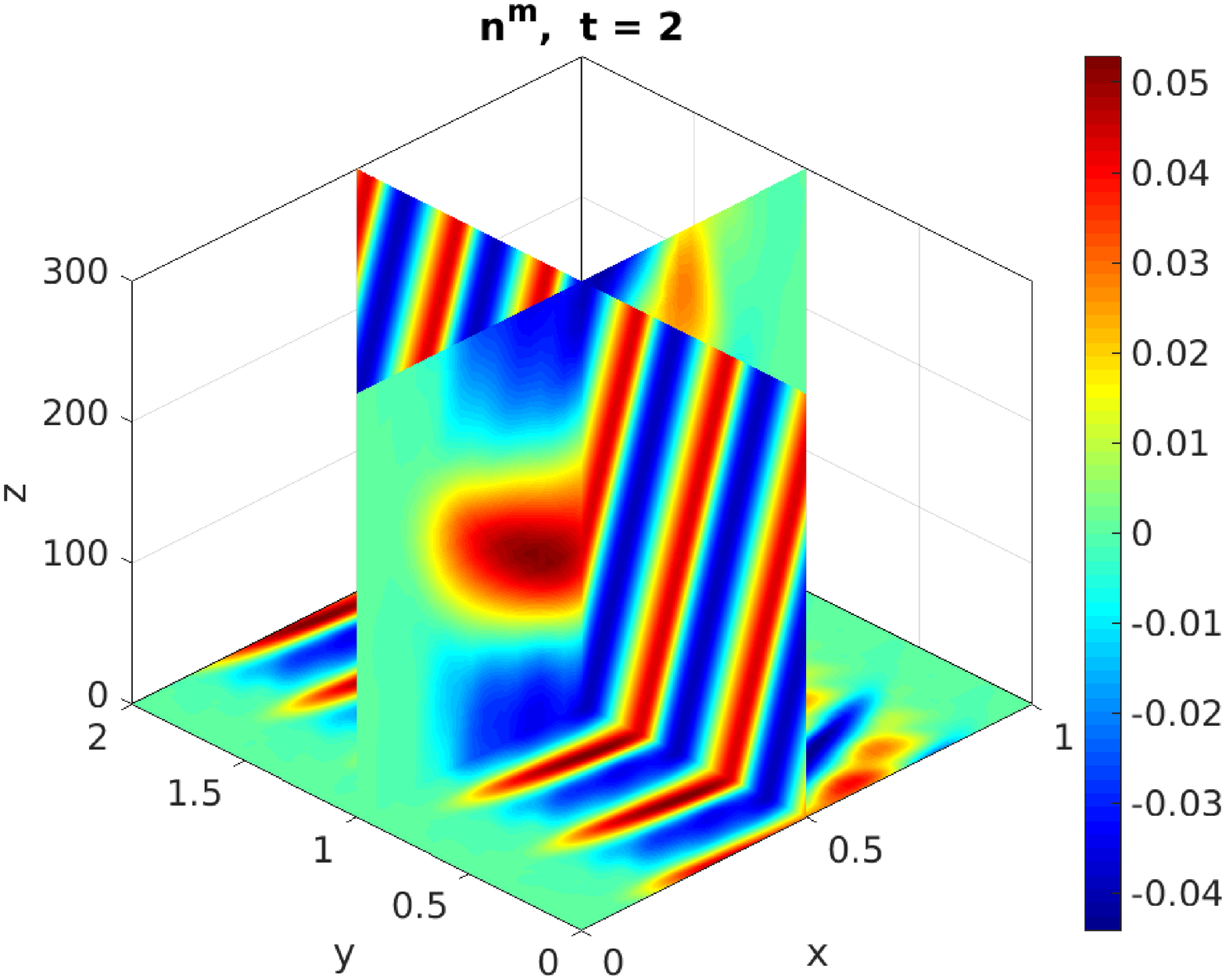}}%
    \hspace*{0.05\textwidth}%
    \subfigure[Inst. growth rate.\label{fig:ITG:c}]{%
      \includegraphics[width=0.3\textwidth]{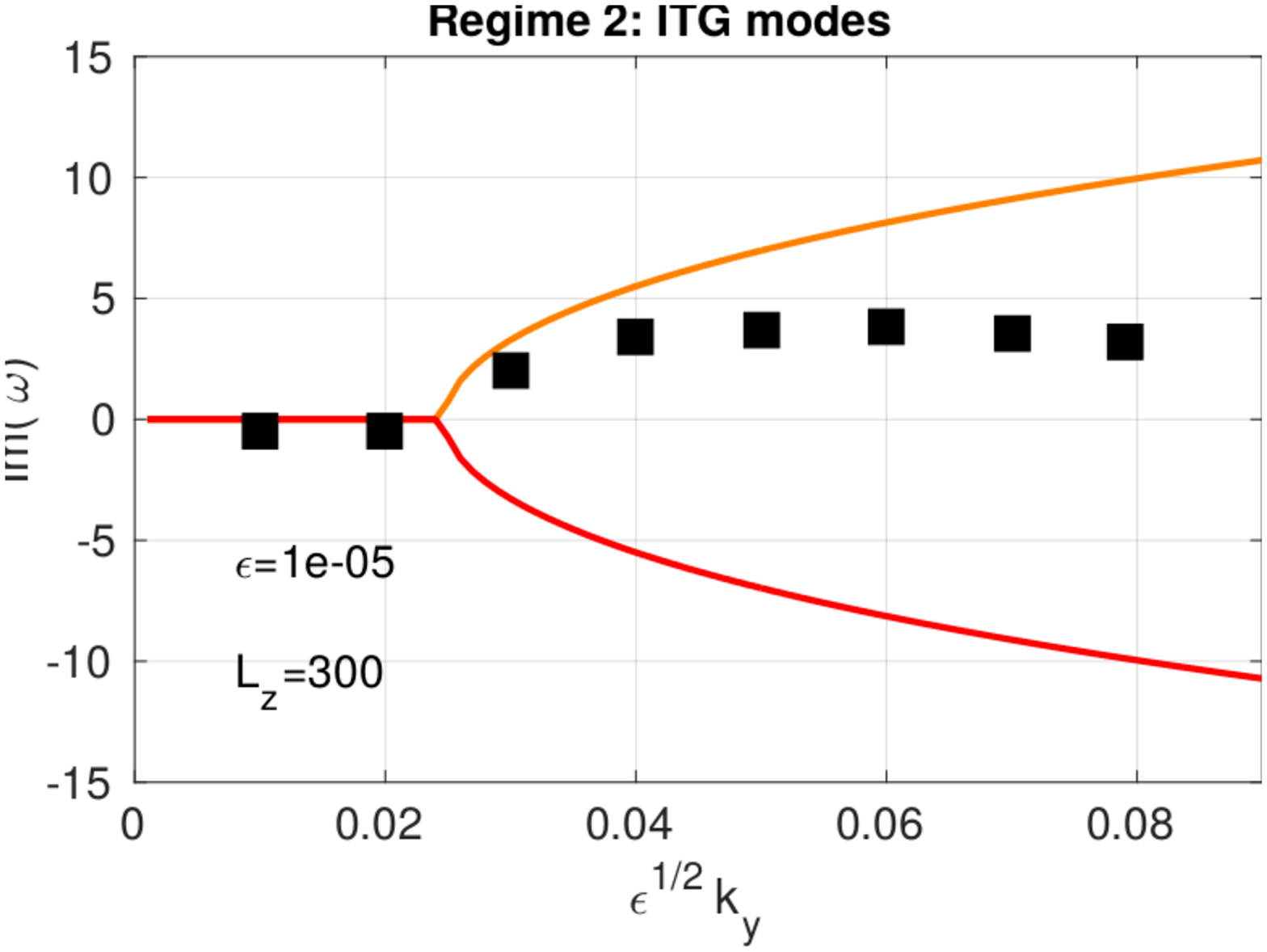}}%
    \abovecaptionskip 1pt
    \caption{Drift-AP scheme simulations: three dimensional slab ITG simulation thanks to AP-scheme implementing the parallel momentum reformulation on a $40\times200\times20$ mesh: density perturbation at times t=1 (a) and t=2(b) and comparison of the ITG growth rate (c) as estimated from the simulation (black square) and the analytic values (solid line). The instability trigger threshold is accurately reproduced, the growth rate being underestimated due to the numerical diffusion of the implementation (see the discussion in \cite{ITGDPO}).}\label{fig:ITG}
  \end{small}
\end{figure}

Finally this approach is used in \cite{ITGDPO} for the bi-fluid plasma description, the electron being assumed in the Boltzmann approximation (section~\ref{sec:EL:Boltzmann}). An energy equation is added to the system describing the ions with gyro-viscous terms from the Braginskii closure \cite{braginskii_transport_1965}. The magnetic field is constant and aligned with one coordinate, periodic boundary conditions being prescribed at each end of the field lines. The anisotropic diffusion equation stemming from the reformulation is solved thanks to the duality based AP method (see section~\ref{sec:AP:Lagrangien}) for the three dimensional simulation of the slab ion temperature gradient instability (ITG \cite{grandgirard_drift-kinetic_2006,jolliet_global_2007,hill_effect_2015,hazeltine_framework_2004}) as depicted by the figure~\ref{fig:ITG}.

\subsection{Pressure based reformulation}\label{sec:ref:pression}\label{sec:EL:ref:P}

Another approach proposed in \cite{degond_numerical_2013} and implemented in \cite{brull_asymptotic-preserving_2011} solves the acoustic equation in order to impose that the force imbalance along the magnetic field lines is bounded by $\tau$, with
\begin{equation}
  \label{eq:condition}
  - \nabla_\parallel p + n E_\parallel = \mathcal{O}(\tau) \,.
\end{equation}
This equilibrium is computed in order to prevent the degeneracy of the parallel momentum equation, with, in the drift limit, the set of equations
\begin{subequations}\label{eq:EL:iso:ion:deriv}
  \begin{align}
    & \frac{\partial n}{\partial t} + \nabla \cdot q = 0 \label{eq:EL:ion:iso:deriv:a} \,, \\
      &q_\perp = \frac{b}{|B|} \times \left( n E + \nabla p\right)\label{eq:EL:ion:iso:deriv:b}\,, \\
    & \frac{\partial}{\partial t} q_\parallel+ \nabla \cdot \left( q \otimes \frac{q}{n} \right)_\parallel =\lim_{\tau \rightarrow 0}  \Big( \frac{1}{\tau}\Big( - T \nabla_\parallel n
   + n E_\parallel\Big)\Big) \,,  \label{eq:EL:ion:iso:deriv:c}  
  \end{align}
\end{subequations}
providing all the quantities, including $q_\parallel$ if the condition \eqref{eq:condition} is met.

This procedure has only been developed is the context of an isothermal plasma description. The set of equations considered does not incorporate the energy equation and the pressure gradient reduces to $\nabla p = T \nabla n$. The reformulation is derived from the momentum equation projected along the magnetic field lines
\begin{equation*}
   \nabla_\parallel \cdot \frac{\partial q_\parallel}{\partial t} = \frac{1}{\tau} \nabla_\parallel \cdot \Big( - T \nabla_\parallel n + n E_\parallel\Big) -  \nabla^2_\parallel : \mathbb{S} \,, \quad   \nabla^2_\parallel : \mathbb{S} := \nabla_\parallel \cdot \left(\nabla \cdot \left( q \otimes \frac{q}{n} \right)_\parallel\right)\,. 
\end{equation*}
together with the density equation, rather than the energy equation (see section~\ref{sec:EL:drift}) yielding to the equation of the acoustic waves for isothermal descriptions
\begin{equation*}
   \tau \frac{\partial^2 n}{\partial t^2} - \nabla_\parallel \cdot ( T \nabla_\parallel n)  = \nabla_\parallel \cdot (n E_\parallel) - \tau \left( \nabla \cdot q_\perp - \nabla^2_\parallel : \mathbb{S}\right) \,.
\end{equation*}
This last equation provides a way to compute the plasma density which secures Condition \eqref{eq:condition}.  This reformulation is implemented and validated in \cite{brull_asymptotic-preserving_2011}. It borrows some of the concepts of low-Mach regime numerical methods (see for instance \cite{degond_all_2011}) that could be transposed to this framework.
\subsection{Self-consistent electric field computation}\label{sec:calcul:E}
\subsubsection{Quasi-neutrality equation}\label{sec:SCE:QN}

The quasi-neutral regime prevails  when the Debye length is small compared to the space scale characterizing the plasma evolution (see section \ref{sec:EL:drift}), which amounts to taking the limit $\lambda \to 0$. In this asymptotic limit, the Poisson equation \eqref{eq:poisson:adim} degenerating into $\rho =0$ cannot provides a means of computing the electric potential. As mentioned in section~\ref{sec:ref:Maxwell}, in this asymptotic, the electric field can be regarded as the Lagrange multiplier of the quasi-neutrality constraint $\nabla \cdot J=0$, $J$ being the current of particles. 

To derive the equation verified by $\phi$ the conservation of the electronic \eqref{eq:EL:elec:Adim:a} and ionic \eqref{eq:EL:ion:Adim:a} densities are combined to provide
\begin{equation}
  \frac{\partial \rho}{\partial t} + \nabla \cdot J = 0\,, 
\end{equation}
with $\rho = n- n_e$ and $J = q - q_e$. The quasi-neutrality constraint is time differentiated, yielding
\begin{equation}
  \nabla \cdot \left(\frac{\partial q}{\partial t} - \frac{\partial q_e}{\partial t} \right) = 0 \,,
\end{equation}
so that the momentum equations can be used to introduce a contribution of the electric field into this relation by means of the electric force definition. This is similar to the reformulation derived in the section \ref{sec:ref:Maxwell}.

This path has been explored in \cite{brull_asymptotic-preserving_2011} with the following time semi discretization 
\begin{subequations}
\begin{align}
  &\frac{1}{\Delta t} \left(n^{m+1} - n^{m} \right) + \nabla_\parallel \cdot q_\parallel^{m+1} + \nabla \cdot q_\perp^m = 0 \,,\label{eq:EL:reform:P:masse}\\
\begin{split}
&\frac{1}{\Delta t} \left(q^{m+1} - q^m \right) + \nabla \cdot \mathbb{S} + \frac{1}{\tau} T\nabla n ^{m+1} = \\
&\hspace*{5.5cm}\frac{1}{\tau} \left( n^{m+1} \nabla \phi^{m+1} + q^{m+1}\times B\right) \,,\label{eq:EL:reform:P:qe}
    \end{split}\\
  \begin{split}
      &\frac{1}{\Delta t} \left(q_e^{m+1} - q_e^m \right) + \nabla \cdot \mathbb{S}_e + \frac{1}{\varepsilon\tau} T\nabla n ^{m+1} =\\
& \hspace*{5.5cm}-\frac{1}{\varepsilon\tau} \left( n^{m+1} \nabla \phi^{m+1} + q_e^{m+1}\times B\right) \,.\label{eq:EL:reform:P:q}
  \end{split}
\end{align}
\end{subequations}
On the discrete level, the quasi-neutrality constraint can be discretized as
\begin{equation*}
  \nabla \cdot (q^{m+1}- q_e^{m+1}) = 0 \,,
\end{equation*}
giving rise to an equation for $\phi^{m+1}$
\begin{equation}\label{eq:phi:illposed}
 -  \nabla_\parallel \cdot \left( n^{m+1}(1+\varepsilon^{-1}) \nabla_\parallel \phi^{m+1}\right) = \tau \mathcal{S} \,.
\end{equation}
We refer to \cite{brull_asymptotic-preserving_2011} for the detailed algebra and the expression of $\mathcal{S}$.

The equation \eqref{eq:phi:illposed} is ill posed, when supplemented by either Neumann or periodic boundary conditions (at the field line extremities) whatever the values of the asymptotic parameter $\tau$. To overcome this difficulty, the problem is regularized in \cite{brull_asymptotic-preserving_2011} adding an extra contribution of the electric potential in the density conservation, with
\begin{equation*}
   \frac{\partial n}{\partial t} + C \frac{\partial \phi}{\partial t} + \nabla \cdot q = 0\,, 
\end{equation*}
$C$ being a small perturbation parameter. The perturbed  system allows the construction of a diffusion equation for $\phi$:
\begin{equation*}
 -  \nabla_\parallel \cdot \left( n^{m+1} \nabla_\parallel \phi^{m+1}\right) + \tau \kappa \phi^{m+1}= \tau \mathcal{S}' \,,
\end{equation*}
 $\kappa$ being a function of the problem data and the numerical parameters. This equation falls within the class of strongly diffusion problem, similar to the one verified by $n$. It is solved by the differential characterization method presented in section~\ref{sec:AP:diff}.

\Fabrice{  \begin{remark}\label{rem:reform:order}
The reformulation consists in making explicit an additional contribution of the parallel velocity in the momentum conservation equation. Manipulating the continuous equations, this algebra requires the time differentiation of the parallel momentum equation in order to insert the expression of the pressure as a function of the momentum. This time derivative is not compulsory in the discrete system. Working the time discretization, it is possible to elaborate a parabolic equation for this quantity securing the parallel equilibrium \eqref{eq:zero:force:b} rather than its time derivative as suggested by the equation  \eqref{eq:zero:force:dtb}. This is achieved by an implicit time discretization of the parallel momentum in Eq.~\eqref{eq:EL:reform:P:masse} and pressure gradient in Eqs.~\eqref{eq:EL:reform:P:qe} and \eqref{eq:EL:reform:P:q}.  This remark echoes that of section \ref{sec:QN:num} (remark~\ref{rem:form:AP}).
 \end{remark}}

\begin{remark}\label{rem:elliptic:aniso}
Another way to render the well posedness of the problem verified by $\phi$ is to compute the perpendicular momentum in \eqref{eq:EL:reform:P:masse} implicitly in order to bring the transverse derivatives back into Eq.~\eqref{eq:phi:illposed}. The equation thus obtained is an anisotropic elliptic equation, whose prototype is investigated in section~\ref{sec:AP:Ellipti:intro}.
\end{remark}

\subsubsection{The Boltzmann relation for the electrons}\label{sec:EL:Boltzmann}

This approximation consists in neglecting the particle inertia in the electronic momentum equation \eqref{eq:EL:elec:Adim}, which amounts to letting $\varepsilon \to 0$, assuming a temperature with vanishing aligned gradients. This last assumption is justified by the heat flux responsible for a rapid balancing of the temperature along the magnetic field lines \cite{braginskii_transport_1965,huba_nrl_2011}. 
With these assumptions, the electronic momentum equation gives rise to the following equilibrium 
\begin{equation*}
  T_e \nabla_\parallel n_e = n_e \nabla_\parallel \phi \,,
\end{equation*}
Assuming the quasi-neutrality of the plasma $n=n_e$, the Maxwell-Boltzmann relation \cite{langmuir_interaction_1929} can be stated 
\begin{equation}\label{eq:elec:Boltzmanien}
  n = n_0 \exp\left(\frac{\phi}{T}\right)\,,
\end{equation}
with $n_0$ an equilibrium density verifying  $\nabla_\parallel n_0 = 0$. This system closure  is implemented in  \cite{ITGDPO} for the self-consistent computation of the electric field.

\section{Numerical methods for strongly anisotropic elliptic and diffusion equations}
\label{sec:aniso}
\subsection{Introduction}\label{sec:AP:Ellipti:intro}

This section is devoted to an overview of methods designed for the numerical resolution of elliptic (or diffusion) equations with large anisotropies. This is a class of problems representative of the difficulty stemming from the simulation of plasmas under a large magnetic field. Tokamak plasmas are a good example of this kind of problems. In this framework the difficulty lies in the periodic boundary conditions applied at the field line extremities in order to account for the periodicity of the torus. Other fields of application can be named, with for instance ionospheric plasma simulation with to so-called Dynamo-3D model \cite{besse_model_2004} presenting the same difficulty, however with Neumann boundary conditions prescribed at each end of the magnetic lines.

The difficulty just mentioned is outlined on a simplified toy model, consisting of an anisotropic elliptic equation posed in a cuboid domain  $\Omega_x\times\Omega_z$, the boundaries being $\Gamma_x = \partial \Omega_x$ and $\Gamma_z = \partial \Omega_z$. The anisotropy strength is denoted~$\varepsilon$, 
\begin{equation}\label{eq:def:pb:mod}
  (P^\varepsilon)\left\{\begin{array}[c]{l}    
    \displaystyle -\frac{\partial}{\partial x} \left(A_\perp \frac{\partial\phi^\varepsilon}{\partial x} \right) - \frac{1}{\varepsilon}  \frac{\partial}{\partial z}  \left( A_\parallel \frac{\partial \phi^\varepsilon}{\partial z} \right) = f^\varepsilon  \,,  \quad \text{ in } \Omega_x \times \Omega_z \,,\\[2mm]
    \displaystyle \frac{\partial \phi^\varepsilon}{\partial z} = 0  \,, \quad  \text{ on } \Gamma_z \,,\\[2mm]
   \displaystyle \phi^\varepsilon = 0 \,, \quad  \text{ on } \Gamma_x \,,
  \end{array}\right.
\end{equation}
$A_\perp$ and $A_\parallel$ being two positive functions. In the Dynamo-3D model mentioned above, the electrostatic potential computed by means of the quasi-neutrality equation, verifies an analogous anisotropic elliptic equation, when the magnetic field is assumed aligned with the $z$ coordinate. A similar equation would be obtained following the remark~\ref{rem:elliptic:aniso} of the section~\ref{sec:SCE:QN}.
The problem associated to the dominant operator in the limit of infinite anisotropy strength is ill-posed, its kernel being populated by the functions that do not depend on the $z$ coordinate. Indeed, multiplying \eqref{eq:def:pb:mod} by $\varepsilon$ and considering formally the limit $\varepsilon \rightarrow 0$ yields
\begin{equation}\label{eq:def:sys:deg}
 (D) \left\{\begin{array}[c]{l}
    \displaystyle \frac{\partial}{\partial z}\left( A_\parallel \frac{\partial \phi^0}{\partial z} \right)= 0 \,,  \quad \text{ in } \Omega_x \times \Omega_z \,,\\[3mm]
    \displaystyle \frac{\partial \phi^0}{\partial z} = 0  \,, \quad \text{ on } \Gamma_z \,.
  \end{array}\right.
\end{equation}
This degenerate system admits an infinite amount of solutions namely all functions $\bar \psi$ only depending on $x$. 
However, $\phi^0$ defined as the limit of $\phi^\varepsilon$, the solution of the problem \eqref{eq:def:pb:mod}, verifies a well posed problem \cite{degond_asymptotic_2010,degond_duality-based_2012}. This system is obtained by integrating the elliptic equation \eqref{eq:def:pb:mod} along the $z$ coordinate. Thanks to the boundary conditions applied on $\Gamma_z$, one can write, in the limit $\varepsilon \rightarrow 0$ 
\begin{equation} \label{eq:def:sys:limit}
  (P^0)\left\{\begin{array}[c]{l}
    \displaystyle - \frac{\partial}{\partial x} \left(\bar A_\perp \frac{\partial\phi^0}{\partial x} \right) = {\bar f}^0 \,,  \quad \text{ in } \Omega_x \,,\\[3mm]
    \displaystyle \phi^0 = 0  \,, \quad \text{ on } \Gamma_x \,.
  \end{array}\right.
\end{equation}
In this equation $\bar f$ is the mean value of $f$ along the  $z$ direction:
\begin{equation}\label{eq:def:moyenne}
  \bar f(x) = \frac{1}{\text{mes}(\Omega_z)}\int_{\Omega_z} f(x,z) \, dz \,,
\end{equation}
and similarly for $\bar A_\perp$.

\begin{figure}[!ht]
  \centering
  \graphicspath{{./FIG/AP/}}
  \begin{minipage}[c]{0.48\textwidth}
    \psfrag{Error}[][][.8]{Condition number}
    \psfrag{Error Norm 2}[][][.8]{}
    \subfigure[Condition number estimate (mesh $50\times 50$).\label{fig:cond:prec:a}]{\includegraphics[width=1.\textwidth]{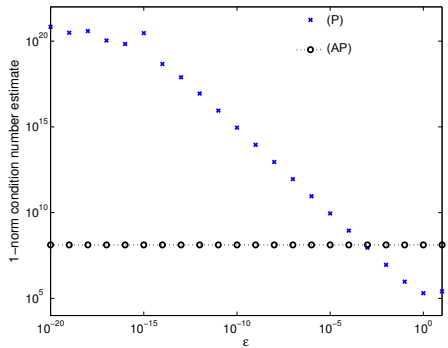}}    
  \end{minipage}\hfill%
  \begin{minipage}[c]{0.48\textwidth}    
    \psfrag{Error}[][][.8]{$l^2$-norm error}
    \psfrag{Error Norm 2}[][][.8]{}
    \subfigure[Norm of the relative error between the exact solution and the numerical approximations (mesh $50\times50$).\label{fig:cond:prec:b}]{\includegraphics[angle=-90,width=1.\textwidth]{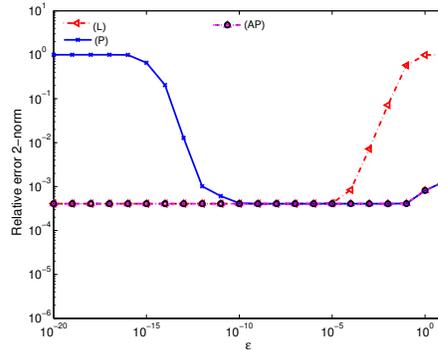}}    
  \end{minipage}\hfill%
  \begin{minipage}[c]{0.48\textwidth}    
    \subfigure[Norm of the relative error between the exact solution and the numerical approximations (mesh $500\times500$).\label{fig:cond:prec:c}]{\includegraphics[angle=-90,width=1.\textwidth]{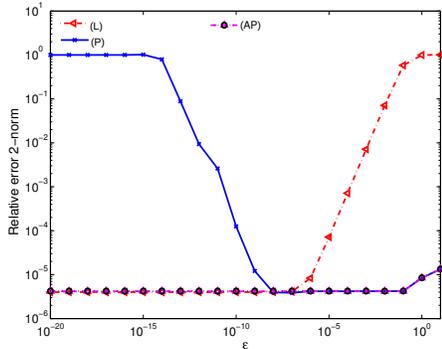}}    
  \end{minipage}\hfill%
  \begin{minipage}[c]{0.48\textwidth} 
  \caption{Property of standard discretizations of the singular perturbation problem \eqref{eq:def:pb:mod} (P) and the limit problem \eqref{eq:def:sys:limit} (L) compared to an AP-method  (AP) : condition number of the linear system obtained thanks to a  $\mathbb{Q}_1$ finite element method, relative error computed between the exact solution and the numerical approximation as functions of the anisotropy strength \protect\cite{degond_asymptotic_2010,besse_efficient_2013}.}
  \label{fig:cond:prec}
\end{minipage}
\end{figure}

The system \eqref{eq:def:sys:limit} is obtained as the limit problem of the anisotropic equation. At this point the singular nature of the problem is clearly outlined. The limit problem \eqref{eq:def:sys:limit} is a one dimensional elliptic problem integrated along the anisotropy direction, while the initial problem \eqref{eq:def:pb:mod} is a two dimensional elliptic equation.
For small values of the asymptotic parameter, standard discretizations of the singular perturbation problem \eqref{eq:def:pb:mod} will become consistent with the degenerate problem \eqref{eq:def:sys:deg}. The conditioning of the system matrix is thus expected to blow up with vanishing  $\varepsilon$-values as reported on the plots of Fig.~\ref{fig:cond:prec:a}.
Jointly with the conditioning blow up, the precision of the numerical approximation cannot be preserved for large anisotropy strength. This loss of accuracy is found for $\varepsilon$ values all the larger that the mesh is more refined, as shown by approximation errors displayed on Figs.~\ref{fig:cond:prec:b} and \ref{fig:cond:prec:c}. 

The principle of AP-Schemes \cite{degond_asymptotic_2010,degond_duality-based_2012,besse_efficient_2013,degond_asymptotic-preserving_2012,lozinski_highly_2012, narski_asymptotic_2013} is to secure the consistency of the discrete system with the limit problem \eqref{eq:def:sys:limit} when $\varepsilon \rightarrow 0$. In order to harness the microscopic information lost in the degenerate problem \eqref{eq:def:sys:deg}, these methods implement a decomposition of the solution into a part belonging to the kernel $\mathcal{G}$ of the dominant operator, supplemented by a correction, giving rise to
\begin{equation*}
  \phi(x,z) =  \bar \phi + \phi'(x,z) \,, \quad \forall (x,z) \in \Omega_x \times \Omega_z \,,
\end{equation*}
where $\bar \phi \in \mathcal{G}$ is the mean of the function $\phi$ as defined by \eqref{eq:def:moyenne} and $\phi'$ is the fluctuating part verifying 
\begin{equation*}
  \bar \phi' = 0 \,.
\end{equation*}
The difficulty lies now in the discretization of the properties verified by these two components. Different approaches have been developed and detailed in the next sections.

\subsection{Duality based reformulation}\label{sec:AP:Lagrangien}
The first implementation of the AP concepts have been carried out in \cite{degond_asymptotic_2010} with a system consisting of an equation for the mean part of the solution coupled to an equation for the fluctuation:
\begin{equation}\label{de:sys:AP:moyenne}
\left\{\begin{array}[c]{l}
   \displaystyle - \frac{\partial }{\partial x} \left( \bar A_\perp \frac{\partial \bar \phi}{\partial x} \right) = \bar f
     + \frac{\partial}{\partial x}\left(\overline{A_\perp'\frac{\partial \phi'}{\partial x}} \right)\,,\quad \text{in } \Omega_x\,, \hspace*{2.9cm}\\[2mm]
   \displaystyle \bar \phi = 0 \,, \quad \text{ on } \Gamma_x \,,   
  \end{array} \right.
\end{equation}
\begin{equation}\label{de:sys:AP:fluctuation}
   \left\{\begin{array}[c]{l}
         \displaystyle - \varepsilon
     \frac{\partial}{\partial x}\left(A_\perp \frac{\partial \phi'}{\partial x}\right)   - \frac{\partial}{\partial z} \left(A_\parallel \frac{\partial \phi'}{\partial z }\right) + \varepsilon
     \frac{\partial}{\partial x} \left(\overline{A_\perp' \frac{\partial \phi'}{\partial x}} \right) = \\[2mm]
     \hspace*{4cm}\displaystyle \varepsilon f' + \varepsilon \frac{\partial}{\partial x}\left({A_\perp'\frac{\partial \bar \phi}{\partial x}} \right)\,, \text{ in }  \Omega_x\times\Omega_z \,,\\[3mm]
         \displaystyle {\partial_z \phi'} = 0 \quad \text{on }
          \Omega_x \times \Gamma_z \,, \quad
          \displaystyle \phi' = 0 \quad \text{on } \Gamma_x
          \times \Omega_z\,, \quad
          \displaystyle \overline{ \phi'}= 0 \quad \text{in } \Omega_x\,. 
          \end{array} \right.
\end{equation}  
In the limit  $\varepsilon \rightarrow 0$ the degenerate problem  \eqref{eq:def:sys:deg} is recovered from the equation \eqref{de:sys:AP:fluctuation}, however this equation is verified by the only fluctuating part rather than the entire solution. The zero mean property verified by the fluctuation restores the well posedness of the system for $\varepsilon=0$ with $\phi'=0$ as unique solution. Inserting this identity into \eqref{de:sys:AP:moyenne}, the limit problem \eqref{eq:def:sys:limit} is recovered. This demonstrates that, the limit $\varepsilon\to 0$ is regular in the reformulated system and consequently, that the formulation (\ref{de:sys:AP:moyenne}-\ref{de:sys:AP:fluctuation}) is Asymptotic-Preserving.

When the anisotropy direction is aligned with one coordinate, the discretization of the functional space containing the mean function is straightforward. A weak formulation of the problem can be stated as 
\begin{equation}\label{eq:def:sys:Lagrangien:bar}
  \begin{split}
   &\text{Find } \bar \phi \in \mathcal{G} := \left\{ \bar \psi \in H^1(\Omega_x) \, | \, \bar \psi = 0 \text{ on } \Gamma_x \right\} \\
   &(A_\perp \partial_x \bar \phi, \partial_x \bar \psi) = (\bar f, \bar \psi) + (\overline{A_\perp' \partial_x \phi'}, \partial_x \bar \psi)\,, \qquad \forall \bar \psi \in \mathcal{G} \,.    
  \end{split}
\end{equation}
with $(\phi,\psi):=\int_{\Omega} \phi \psi \, dx dz$. The task is more intricate for the discretization of the functional space $\mathcal{A}$ populated by the fluctuations. Introducing  
\begin{equation*}
  \mathcal{V} := \left\{ \psi' \in H^1(\Omega_x\times\Omega_z)\, | \, \psi'=0 \text{ on } \Gamma_x\times\Omega_z \right\} \,,
\end{equation*}
a weak formulation of the problem \eqref{de:sys:AP:fluctuation} providing $\phi'$  is
\begin{gather*}
  \text{Find }  \phi' \in \mathcal{A} := \left\{ \psi'\in \mathcal{V} \, | \, \bar \psi'=0 \right\} \,,\notag \\
( A_\parallel \partial_z \phi', \partial_z \psi') + \varepsilon ( A_\perp \partial_x \phi' ,\partial_x \psi') - \varepsilon( \overline{A_\perp \partial_x \phi'}, \partial_x \psi') = \hspace*{3cm}\\
\hspace*{5cm}\varepsilon (f,\psi') - \varepsilon (A_\perp' \partial_x\bar \phi,\partial_x\psi') \,,  \quad \forall \psi' \in \mathcal{A}\,.\notag
\end{gather*}

\begin{figure}[!ht]
\belowcaptionskip -0.5cm
  \begin{center}
    \graphicspath{{./FIG/AP-Mat/}}
      \subfigure[$\mathcal{M}_1=A$ \label{figmatrixdynamo}]{%
      \begin{minipage}[c]{0.2\textwidth}
        \begin{minipage}[t]{\textwidth}
          \includegraphics[width=0.8\textwidth]{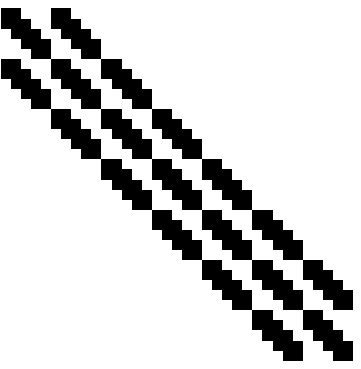}
          \vspace*{0.2\textwidth}
        \end{minipage}%
      \end{minipage}}%
      \hspace*{0.066\textwidth}\subfigure[$\mathcal{M}_2$\label{figmatrixAP}]{%
      \begin{minipage}[c]{0.2\textwidth} 
        \begin{minipage}[t]{\textwidth}        
          \includegraphics[width=0.9\textwidth]{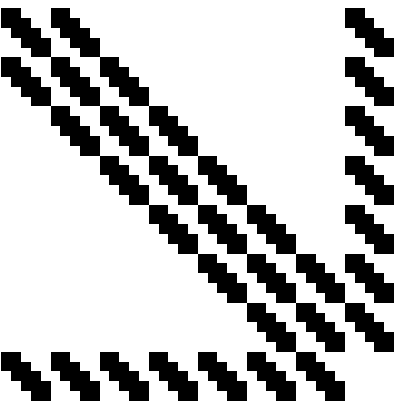}
          \vspace*{0.1\textwidth}          
        \end{minipage}
      \end{minipage}}%
    \hspace*{0.066\textwidth}\subfigure[$\mathcal{M}_3$\label{figmatrixAPM}]{%
      \begin{minipage}[c]{0.2\textwidth} 
        \begin{minipage}[t]{\textwidth}        
          \includegraphics[angle=-90,width=\textwidth]{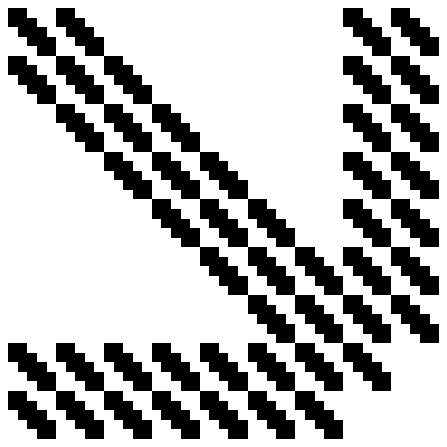}
        \end{minipage}
      \end{minipage}}%
    \hspace*{0.066\textwidth}\subfigure[$\mathcal{M}_O$\label{figmatrixAPO}]{%
      \begin{minipage}[c]{0.2\textwidth} 
        \begin{minipage}[t]{\textwidth}        
          \includegraphics[width=0.9\textwidth]{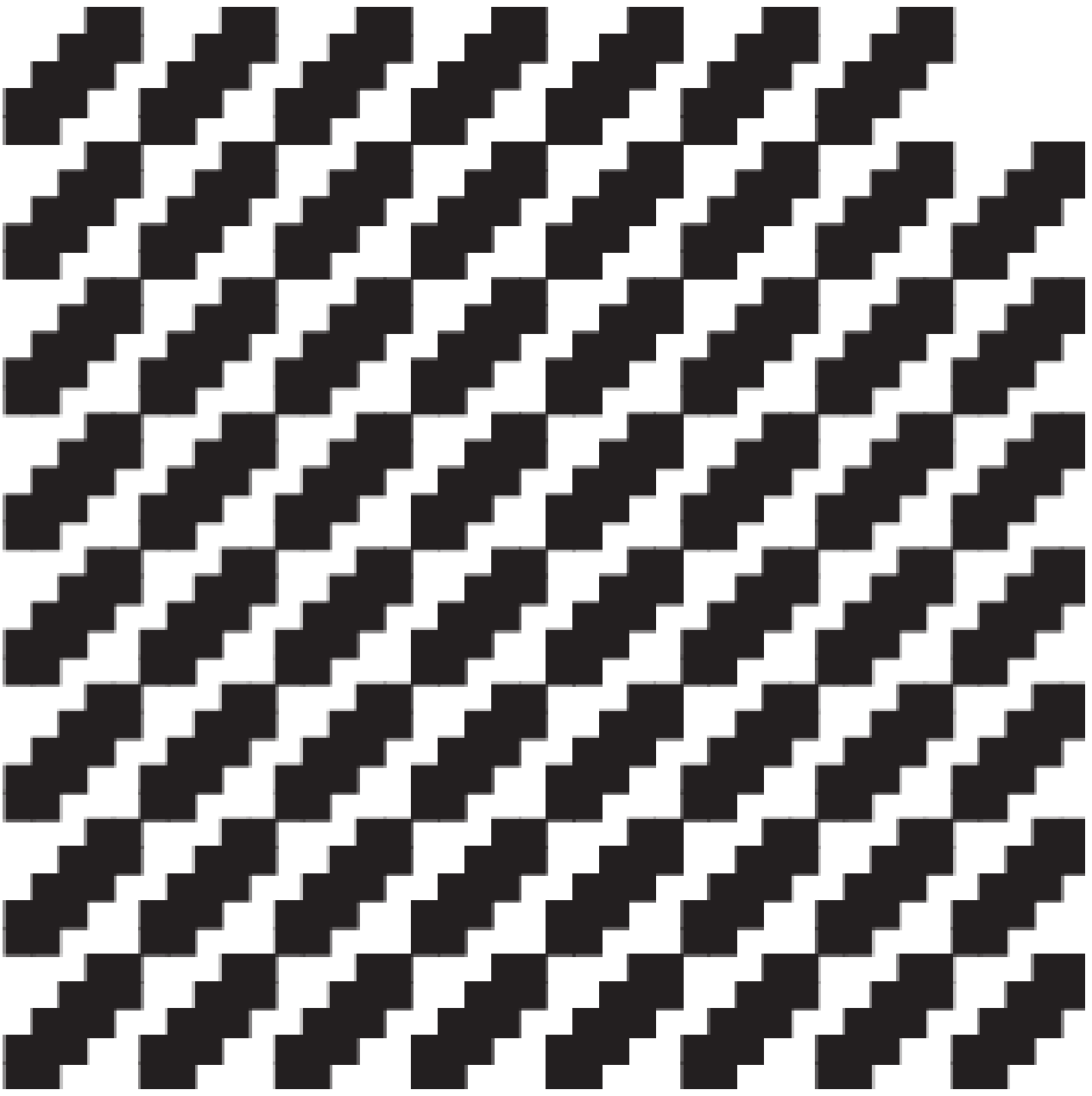}
          \vspace*{0.1\textwidth}          
        \end{minipage}
      \end{minipage}}%
    \begin{small}
\begin{tabular}{|c|c|c|}
\hline
Mat.&$\mathcal{M}_1=A$&$\mathcal{M}_2={\scriptsize
  \left(\begin{array}[c]{cc}
    A&B\\ B^T &0
  \end{array}\right)}$\\
\hline
\# rows &$N_x(N_z+2)$&$N_x(N_z+3)$\\
\hline
Nnz &$(3N_z+4)(3N_x-2)$&$(5N_z+8)(3N_x-2)$ \\
\hline
\hline
Mat.&$\mathcal{M}_3={\scriptsize\left(\begin{array}[c]{ccc}
    A &\varepsilon C &B\\ \varepsilon C^T &\varepsilon A_2& 0 \\ B^T & 0 & 0
  \end{array}\right)}$& $\mathcal{M}_O={\scriptsize
  \left(\begin{array}[c]{cc}
    \bar{A} &B\\ B^T &0
  \end{array}\right)}$\\
\hline
\# rows &$N_x(N_z+4)$&$N_x(N_z+3)$\\
\hline
Nnz &$(7N_z+13)(3N_x-2)$&$(N^2_z+6N_z+8)(3N_x-2)$ \\
\hline
\end{tabular}      
\end{small}
\caption{Matrices obtained thanks to a $\mathbb{Q}_1$-FEM discretization of (a) the singular perturbation problem \eqref{eq:def:pb:mod}, (b) the fluctuation equation~\eqref{de:sys:AP:fluctuation:bis}, (c)  the AP formulation (\ref{de:sys:AP:moyenne}-\ref{de:sys:AP:fluctuation:bis}), (d) the original fluctuation equation~\eqref{de:sys:AP:fluctuation}: plots of the structure pattern for a grid $(N_x,N_z)=(5,5)$ cells, matrix number of rows (\# rows) and number of non zeros elements (Nnz).}\label{fig:matrices:structure}
\end{center}
\end{figure}

The discretization of $\mathcal{V}$ is avoided by the introduction of a Lagrangian aimed at penalizing the zero mean constraint verified by the fluctuations. The weak formulation of the problem can thus be recast into
\begin{equation}\label{eq:def:sys:Lagrangien:prime}
  \begin{split}
  &\text{Find } \phi' \in \mathcal{V} \text{ and }  \bar P \in \mathcal{G} \text{ such that}  \\
&\left\{\begin{array}[c]{l}
    \displaystyle ( A_\parallel \partial_z \phi', \partial_z \psi) + \varepsilon ( A_\perp \partial_x \phi' ,\partial_x \psi) - \varepsilon( \overline{A_\perp \partial_x \phi'}, \partial_x \psi) + (\bar P, \psi)= \\[3mm]
\displaystyle \hspace*{4.5cm}\varepsilon (f,\psi) - \varepsilon (A_\perp' \partial_x\bar \phi,\partial_x\psi) \,,  \quad \forall \psi \in \mathcal{V}\,, \\[3mm]
\displaystyle (\bar \chi, \phi') = 0 \,,\qquad  \forall \bar \chi \in \mathcal{G}
\end{array}\right.    
  \end{split}
\end{equation}
 The equivalence of the reformulated system~(\ref{eq:def:sys:Lagrangien:prime}-\ref{eq:def:sys:Lagrangien:bar})  with the singular perturbation problem \eqref{eq:def:pb:mod} is demonstrated in \cite{degond_asymptotic_2010}.
A finite element discretization gives rise to an augmented linear system with the matrix denoted by $\mathcal{M}_O$,
\begin{equation*}
  \mathcal{M}_O \left(\begin{array}[c]{c}
      \Phi_h \\
      \bar P_h
    \end{array} \right)= \left(\begin{array}[c]{c}
      F_h \\
      0
    \end{array} \right) \,, \qquad \mathcal{M}_O = \left(
    \begin{array}[c]{cc}
      \bar{A} & B\\
      B^T & 0
    \end{array}
\right)  \,,
\end{equation*}
This matrix sparsity pattern is represented on Fig.~\ref{fig:matrices:structure}. If $ N_x\times N_z$ denotes the number of cells of the mesh, the block-matrix $\bar A$ denotes the finite element discretization of the integro-differential operator applied to $\phi'$ on the left hand side of the equation \eqref{eq:def:sys:Lagrangien:prime},  $\bar A \in \mathbb{R}^{N_x(N_z+2)\times N_x(N_z+2)}$, $B \in \mathbb{R}^{N_x\times N_x(N_z+2)}$ being the discretization of the Lagrangian contribution in the system matrix, $(\Phi_h,F_h) \in \mathbb{R}^{N_x(N_z+2)} \times \mathbb{R}^{N_x(N_z+2)}$ and $P_h \in \mathbb{R}^{N_x}$ denoting the vectors associated to $\phi'$, the right hand side of the system and to the Lagrangian $\bar P$. 

The integral discretization in the fluctuation equation induces a fill-in of the system matrix, with a negative impact on the numerical method efficiency with respect to the memory requirements and the computational time. To improve the efficiency of the method a second reformulation is proposed in \cite{besse_efficient_2013} in which the equation \eqref{de:sys:AP:fluctuation} is substituted by  
\begin{equation}\label{de:sys:AP:fluctuation:bis}
   \left\{\begin{array}[c]{l}
         \displaystyle - \varepsilon
     \frac{\partial}{\partial x}\left(A_\perp \frac{\partial \phi'}{\partial x}\right)  - \frac{\partial}{\partial z} \left(A_\parallel \frac{\partial \phi'}{\partial z }\right)  = \varepsilon f + \varepsilon \frac{\partial}{\partial x}\left({A_\perp\frac{\partial \bar \phi}{\partial x}} \right)\,, 
     \\[3mm]
         \displaystyle {\partial_z \phi'} = 0 \quad \text{on }
          \Omega_x \times \Gamma_z \,, \\[1mm]
          \displaystyle \phi' = 0 \quad \text{on } \Gamma_x
          \times \Omega_z\,, \quad
          \displaystyle \overline{ \phi'}= 0 \quad \text{in } \Omega_x\,.            
          \end{array} \right.
\end{equation}  
The matrix obtained after a FEM discretization of the system (\ref{de:sys:AP:moyenne} - \ref{de:sys:AP:fluctuation:bis}) gives rise to the linear system
  \begin{equation*}
  \left(
    \begin{array}[c]{ccc}
      A & \varepsilon C & B \\
      \varepsilon C^T&  \varepsilon A_2 & 0 \\
      B^T & 0 & 0
    \end{array}
\right) \left(\begin{array}[c]{c}
      \Phi_h \\
      \bar \Phi_h \\
      \bar P_h
    \end{array}
\right) =  \left(\begin{array}[c]{c}
      F_1 \\
      F_2 \\
      0
    \end{array}
\right) \,,
\end{equation*}
 $A \in \mathbb{R}^{N_x(N_z+2)\times N_x(N_z+2)}$ being the matrix discretizing the singular perturbation problem \eqref{eq:def:pb:mod}, $A_2 \in \mathbb{R}^{N_x\times N_x}$ the one obtained after the discretization of the mean part equation \eqref{eq:def:moyenne}, $C \in \mathbb{R}^{N_x\times N_x(N_z+2)}$ being the coupling term with the fluctuation. Finally $\bar \Phi_h \in \mathbb{R}^{N_x}$ is the vector associated with the mean part, $(F_1,F_2)\in\mathbb{R}^{N_x(N_z+2)} \times \mathbb{R}^{N_x}$ define the right hand side of the system.
The plots of the figure~\ref{fig:matrices:structure} show the benefits of this modified formulation, in which the equation for the fluctuation $\phi'$ does not involve any integral operator. On a $500\times 500$ mesh the number of non zeros elements stored in the matrix discretizing the equation \eqref{de:sys:AP:fluctuation} is 168 times larger than that of Problem \eqref{eq:def:pb:mod}, this ratio increasing further with mesh sizes \cite{besse_efficient_2013}.  In contrast, for Problem~(\ref{de:sys:AP:moyenne}-\ref{de:sys:AP:fluctuation:bis}) the non zeros elements remain 2.3 times larger that of Problem \eqref{eq:def:pb:mod} whatever the mesh size.

The efficiency is further improved  by implementing a hybrid method coupling the AP reformulation (\ref{de:sys:AP:moyenne}-\ref{de:sys:AP:fluctuation:bis}) and the limit problem in the region where the asymptotic parameter is small \cite{degond_hybrid_2011,CDN14}. Indeed this last model furnishes a solution that does not depend on the $z$ coordinate, the discretization of this model gives rise to a smaller system matrix. The  hybrid method is thus all the more efficient than the sub-domain on which the limit model is used, can be enlarged. However, as shown by the results gathered in Figs.~\ref{fig:cond:prec:b} and \ref{fig:cond:prec:c}, the limit problem furnishes an accurate approximation only for small values of $\varepsilon$. The modelling error should be comparable to the discretization error in order for the Hybrid-method to provide an approximation as accurate as the AP-method but with a reduced computational cost. 
Table \ref{tab:time8-10} provides some elements relating the gain brought by this coupling strategy.
\begin{table}[!ht]
\begin{center}
\caption{Efficiency of the hybrid (HYB.) and the Asymptotic-Preserving (AP) methods compared to a FEM discretization of the problem~\eqref{eq:def:pb:mod} (P): Number of cells of the grid ($N=N_x= N_z$), resolution time (with the sparse direct solver MUMPS~\cite{amestoy_fully_2001}) relative to that of the singular perturbation problem~\eqref{eq:def:pb:mod} (T), number of non zero elements stored in the factorized matrix (Nnz fact.), number of rows (\# rows), number of non zero elements stored in the system matrix (Nnz Mat.) and precision of the numerical approximation (L$^2$-error norm of the relative error), for $\Omega_z = \Omega^1_z \cup \Omega^2_z$, where $\text{mes}(\Omega^2_z)= \nicefrac{7}{10}\ \text{mes}(\Omega_z)$, $ \Omega^2_z$ being the sub-domain of the limit problem \cite{CDN14}.}
\label{tab:time8-10}
\subfigcapskip 5pt
\subfigbottomskip 10pt
\begin{tabular}{|c|c|r|r|r|r|r|}
\hline
Met. 
& $N$ & \multicolumn{1}{c|}{T} & \multicolumn{1}{c|}{Nnz (fact.)} & \multicolumn{1}{c|}{\#rows} & \multicolumn{1}{c|}{Nnz (Mat.)} & \multicolumn{1}{c|}{error (L$^2$)} \\
\hline
\hline
 (HYB.) 
 & 500 & 53\% &9 861 960 & 77 000 & 1 592 374 & $1.4\times 10^{-5}$ \\
 \hline
 (AP) 
 & 500 & 203\% & 40 655 248 & 252 000 & 5 262 474 & $1.4\times 10^{-5}$ \\
 \hline
 (P) 
 & 500 & 100\% & 26 940 422 & 251 000 & 2 252 992 & $3.4\times 10^{-3}$ \\
 \hline
 \hline
(HYB.) 
& 2000 & 26\% & 206 531 976 & 1 208 000 & 25 269 574 & $1.2\times 10^{-6}$ \\
\hline
(AP) 
& 2000 & 137\% & 804 867 106 & 4 008 000 & 84 049 974 & $8.8\times 10^{-7}$ \\
\hline
(P) 
& 2000 & 100\%  & 557 859 738 & 4 004 000 & 36 011 992 & $1.5\times 10^{-2}$ \\
\hline
\end{tabular}
\end{center}
\end{table}
As mentioned above, the overhead of the AP method compared to the numerical resolution of the singular perturbation problem, with respect to the memory requirements is roughly 2.3 for the most refined meshes. The fill-in of the factorized matrix does not scale as badly, with an increase of the number of non zero elements stored from 50\% to 80\%. When the limit problem is discretized on a large enough sub-domain, the fill-in of the hybrid method is lower than that of the singular problem~\eqref{eq:def:pb:mod} with a computational time necessary for the linear system resolution being divided by 3.

\def\q{\mathfrak{q}}

\subsection{A differential characterization }\label{sec:AP:diff}

\begin{figure}[!ht]
   \centering
     \centering \includegraphics[width=0.6\textwidth]{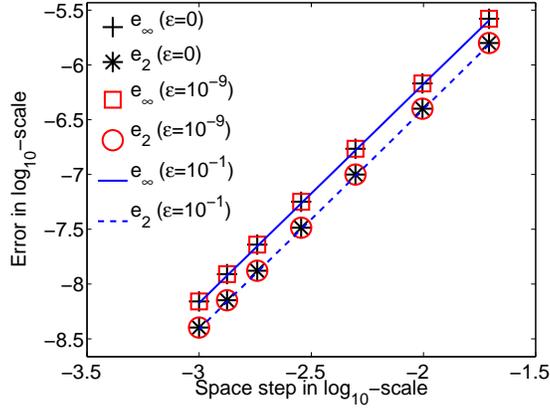}
  \caption{Relative error ($L^2$ and $L^\infty$ norms) between the exact solution and the numerical approximation carried out thanks to differential characterization of the fluctuation space, as a function of the mesh size and, for different anisotropy strengths \protect\cite{brull_numerical_2012}.}\label{cv_inh_linear}
\end{figure}

Another approach is proposed in \cite{brull_numerical_2012} implementing the following characterization of the fluctuation space 
 \begin{equation}
   \begin{split}
   \psi' \in \mathcal{A} \Longleftrightarrow \exists \chi \in \mathcal{W} \,| \,\psi' = \partial_z \chi \quad  \text{ with }\\
   \mathcal{W} := \left\{\eta \in L^2 (\Omega)\, | \,\partial^2_{zz} \eta \in  L^2 (\Omega), \partial^2_{xz} \eta \in  L^2 (\Omega)\,, \eta = 0 \text{ on } \Gamma \right\}\,,     
   \end{split}
\end{equation}
the equation \eqref{de:sys:AP:fluctuation:bis} can thus be recast into
\begin{gather*}
  \text{Find }  \chi \in \mathcal{W} \text{ such that } \notag \\
( A_\parallel \partial_{zz}^2 \chi, \partial_{zz}^2 \eta) + \varepsilon ( A_\perp \partial_{xz}^2 \chi ,\partial_{xz}^2 \eta) = \varepsilon (f,\partial_z \eta) - \varepsilon (A_\perp \partial_x\bar \phi,\partial_{xz}^2\eta) \,,  \quad \forall \eta \in \mathcal{W}\,,\notag
\end{gather*}
For homogeneous coefficients $A_\perp$, a strong formulation of the problem can be stated as
\begin{equation}\label{eq:fourth:order}\left\{
  \begin{array}[c]{ll}
    \displaystyle -\frac{\partial^2 }{\partial z^2} \left(A_\parallel \frac{\partial^2 \chi}{\partial z^2}\right) - \varepsilon A_\perp \frac{\partial^2 }{\partial x^2 } \left( \frac{\partial^2 \chi}{\partial z^2}\right) = \varepsilon \frac{\partial {\widetilde f}}{\partial z}\,, \quad &\text{in } \Omega\,, \\[3mm]
    \displaystyle \chi = 0\,, \quad &\text{on } \Gamma \,,
  \end{array}\right.
\end{equation}
with ${\widetilde f} = f + \frac{\partial }{\partial x} \left( A_\perp \frac{\partial \bar \phi}{\partial x}\right)$.
This system is transformed into two nested elliptic problems for $\phi' = \partial \xi/\partial z$
\begin{subequations}
   \begin{eqnarray}
&    \displaystyle -\frac{\partial^2 }{\partial z^2} \left( A_\parallel  \zeta \right) - \varepsilon A_\perp \frac{\partial^2 \zeta}{\partial x^2 } = \varepsilon \frac{\partial {\widetilde f}}{\partial z}\,, \quad \text{ in } \Omega \,, \qquad
& \displaystyle \zeta = 0 \,, \quad \text{ on } \Gamma\,\label{eq:AP:diff:b}\\[1mm]
&    \displaystyle -\frac{\partial^2 \xi}{\partial z^2} = -\zeta \,, \quad \text{ in } \Omega\,, \qquad&
    \displaystyle \xi = 0 \,, \quad \text{ on } \Gamma_z \,. \qquad \label{eq:AP:diff:c}
  \end{eqnarray}
\end{subequations}
The system  \eqref{eq:AP:diff:c} is well posed and does not depend on $\varepsilon$. Moreover, in the problem \eqref{eq:AP:diff:b}, the dominant operator in the limit $\varepsilon\to 0$ is supplemented with Dirichlet boundary conditions. Its kernel is thus reduced to zero. These two properties define a well posed  problem for all $\varepsilon$. The implementation realized in \cite{brull_numerical_2012} for anisotropic diffusion equations (with $A_\perp=0$) show that the numerical method is AP (see figure~\ref{cv_inh_linear}) providing computations with a precision independent of $\varepsilon$.

The advantage of this approach is to allow the resolution of the anisotropic problem by means of three standard elliptic problems, two nested elliptic problems for the fluctuation and one for the mean part, for which very efficient and proven methods exist.
The recast of the fourth order problem \eqref{eq:fourth:order} into two nested elliptic problems is straightforward for Neumann boundary conditions. However, the generalization to other kinds of boundary conditions and non homogeneous perpendicular coefficients remains to be done.

\subsection{Generalization to non adapted coordinates}\label{sec:extension}

The generalization of the methods introduced in the preceding sections to anisotropy directions non aligned with one coordinate is documented in this section. The singular perturbation problem \eqref{eq:def:pb:mod} is recast into 
\begin{equation}\label{SPprime}
\left\{\begin{array}[c]{ll}
  -\displaystyle \nabla_\perp \cdot \left(A_\perp \nabla \phi^\varepsilon \right) - \frac{1}{\varepsilon} \nabla_\parallel\cdot \left( A_\parallel \nabla_\parallel \phi^\varepsilon \right) =  f^\varepsilon \,,  \quad &\text{ in } \Omega \,,\\[2mm]
  \displaystyle \frac{1}{\varepsilon}n_\parallel \cdot \nabla_\parallel \phi^\varepsilon + n_\perp \cdot \nabla_\perp \phi^\varepsilon= 0 \,, \quad &\text{ on } \Gamma_N \,,\\[2mm]
   \phi^\varepsilon = 0 \,, \quad& \text{ on }  \Gamma_D \,.
\end{array}  \right.
\end{equation}
with for $v\in \mathbb{R}^3$, $v_\parallel := (b\cdot v) b$, $v_\perp :=  (\mathbb{I}d - b \otimes b) v$, and $\nabla_\parallel \cdot v :=  \nabla\cdot v_\parallel \,,\nabla_\perp \cdot v := \nabla \cdot v_\perp \,,$ where $b$ is the unit vector pointing in the direction of the magnetic field $b=B/|B|$ (see equation \eqref{eq:notations:B}). The domain boundary $\Gamma$ is decomposed into $\Gamma=\Gamma_N\cup \Gamma_D$ with $\Gamma_D = \{x\in \Gamma,\, b(x)\cdot n(x) = 0 \}$, $n(x)$ being the outward normal.

\begin{figure}
  \centering
  \subfigure[$\left\|p_{\varepsilon}-p_{\varepsilon,app}\right\|_{\ell^{p}} / \left\|p_{\varepsilon}\right\|_{\ell^{p}}$, $\varepsilon = 10^{-3}$. \label{angle_GH_variable:error_10m3}]{\begin{minipage}[c]{0.49\textwidth}\centering
      \psfrag{Error on}[][][1.]{}
     \includegraphics[width=\textwidth]{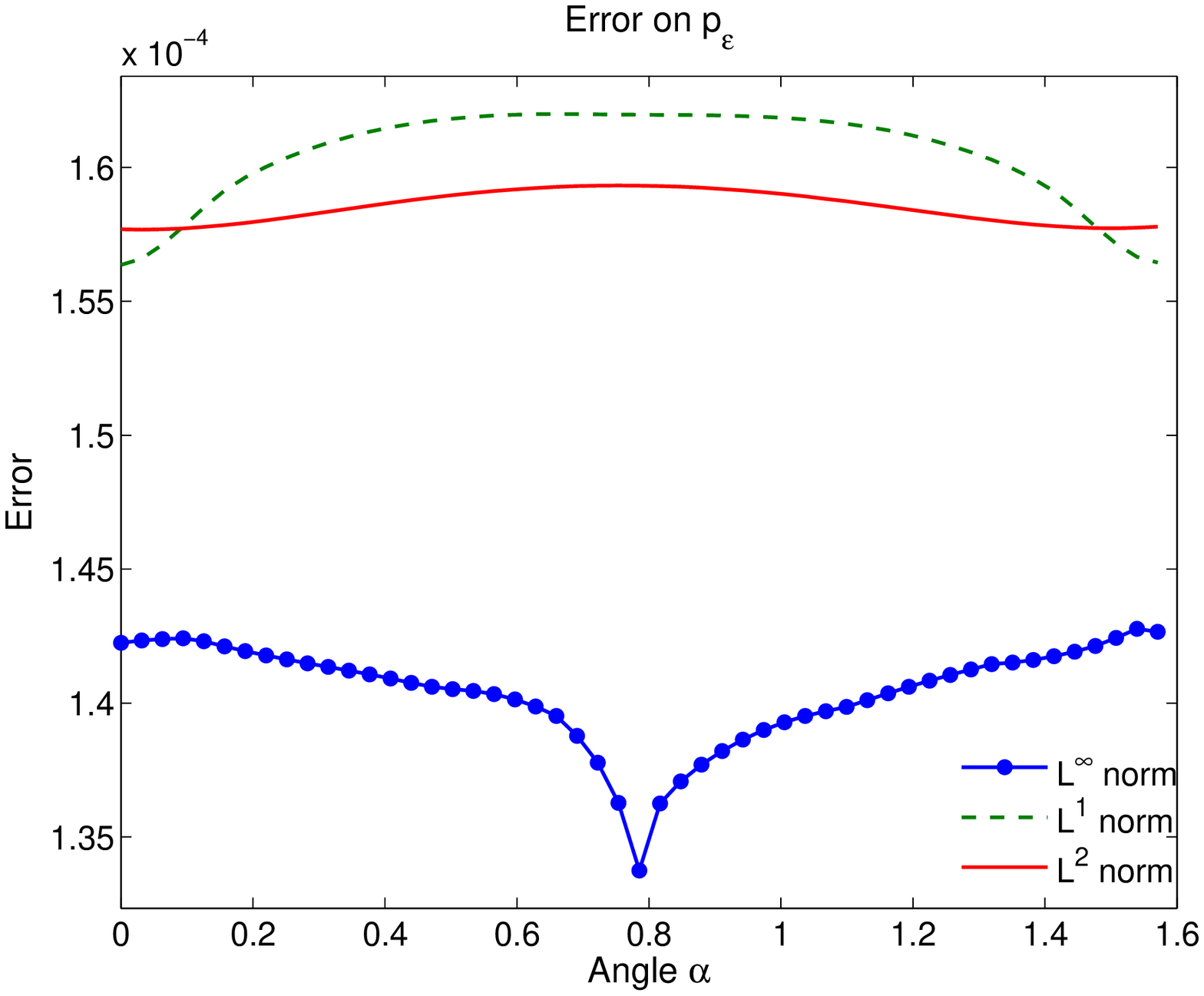}
    \end{minipage}}\hfill%
  \subfigure[$\left\|p_{\varepsilon}-p_{\varepsilon,app}\right\|_{\ell^{p}} / \left\|p_{\varepsilon}\right\|_{\ell^{p}}$, $\varepsilon = 10^{-8}$. \label{angle_GH_variable:error_10m8}]{\begin{minipage}[c]{0.49\textwidth}\centering
      \includegraphics[width=\textwidth]{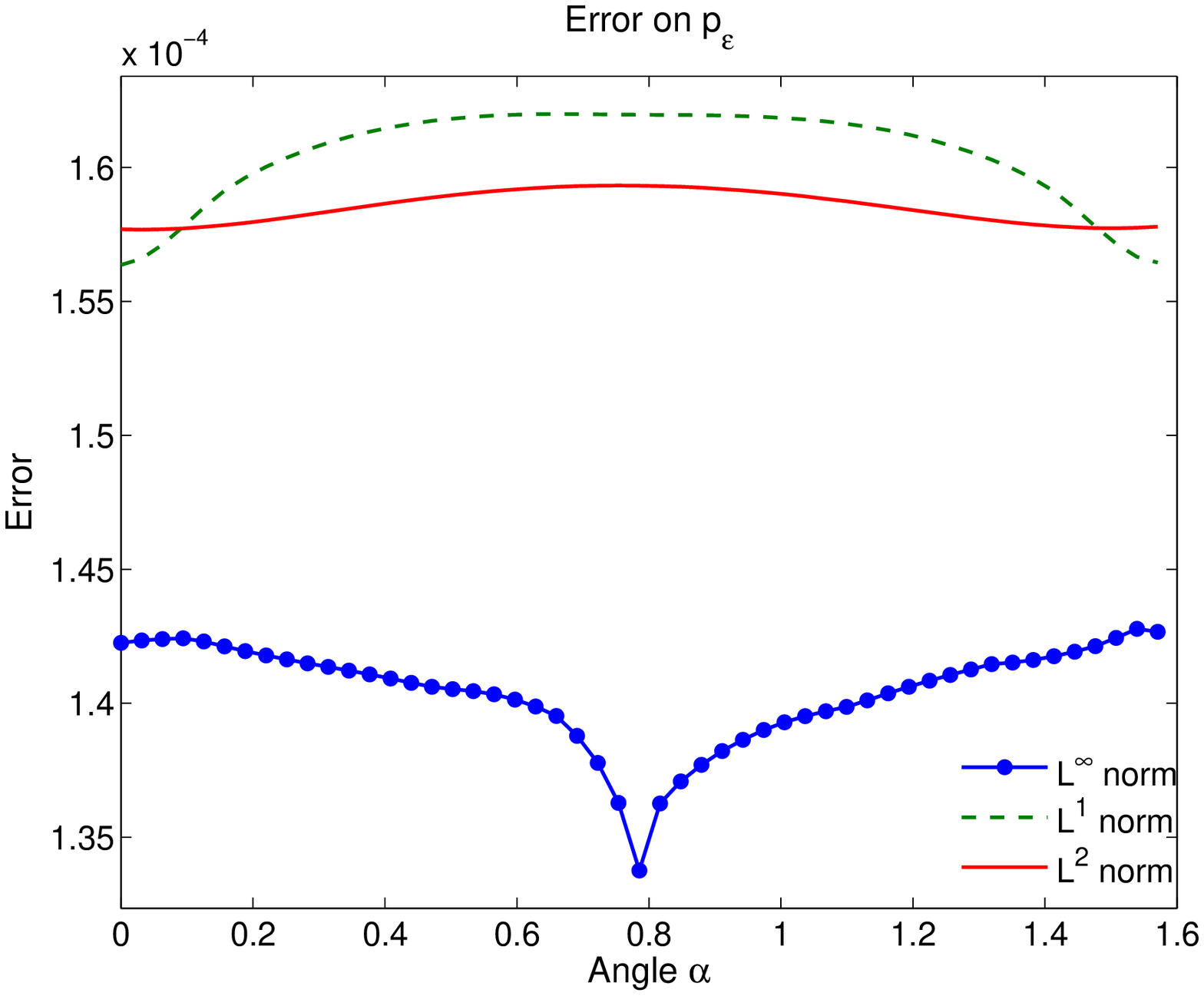}
    \end{minipage}}
  \caption{Numerical implementation of the differential characterization with anisotropies non aligned with one coordinate: Relative error between the exact solution $p_{\varepsilon}$ and the numerical approximation $p_{\varepsilon,app}$ as a function of the $\alpha$ the angle measured between the magnetic field direction and the $x$-axis for different anisotropy strength $\varepsilon = 10^{-3}$ (left) and $\varepsilon = 10^{-8}$ (right). \label{angle_GH_variable}}
\end{figure}

\medskip 
The method implementing the differential characterization is straightforwardly extended to this framework. This approach has been genuinely developed for non adapted coordinates \cite{brull_asymptotic-preserving_2011,brull_degenerate_2012}. More details are available in \cite{brull_numerical_2012}.
The plots of Fig.~\ref{angle_GH_variable} report the precision of the numerical method as a function of the parameter $\alpha$ defined as the angle of the vector $b$ with the horizontal axis. From these computations, the precision of the method is observed to be almost independent of the anisotropy orientation, with a variation in the error norm lower than 10\%.

\medskip

The extension of the duality based formulation is more intricate. Indeed, when one of the coordinates is not aligned with the anisotropy direction, the functional space for the mean value along the anisotropy direction is not easily discretized. The first generalization proposed in \cite{degond_duality-based_2012} implements a computation of the mean by duality. Introducing 
$  \mathcal{V} := \left\{ \phi \in H^1(\Omega)\, | \, \phi = 0 \text{ on } \Gamma_D\right\}$ and the space of the mean functions $
    \mathcal{G} := \big\{ \psi \in \mathcal{V} \, |\, \nabla_\parallel \psi = 0\big\}$,  the space $\mathcal{A}$ of the fluctuations is defined by duality
 \begin{equation*}
  \mathcal{A} := \big\{\varphi \in \mathcal{V}\, | \, (\varphi,\psi)=0 \,, \quad \forall \psi \in \mathcal{G} \big\} \,,  
\end{equation*}
using the orthogonal decomposition  $\mathcal{V} = \mathcal{G}\otimes^\perp \mathcal{A} $, the $L^2$-scalar product being $(f,g) := \int_\Omega f g \,dx \,. $

Therefore, the weak formulation of the singular perturbation problem \eqref{SPprime}
\begin{equation*}
  \begin{split}
  &\text{Find } (p^\varepsilon,q^\varepsilon) \in \mathcal{G}\times\mathcal{A}\,, \text{ such that }  \\
  &\left\{
  \begin{array}[c]{ll}
    (A_\perp \nabla_\perp p^\varepsilon, \nabla_\perp \eta) + (A_\perp \nabla_\perp q^\varepsilon, \nabla_\perp \eta)  = (f,\eta) \,, \, &\forall \eta \in \mathcal{G} \,, \\[2mm]
  (A_\parallel \nabla_\parallel q^\varepsilon, \nabla_\parallel \xi) + \varepsilon (A_\perp \nabla_\perp (q^\varepsilon+p^\varepsilon), \nabla_\perp \xi)  = (f,\xi) \,, \qquad &\forall \xi \in \mathcal{A} \, \\
  \end{array}\right.    
  \end{split}
\end{equation*}
can be substituted by 
\begin{equation*}
  \begin{split}
  &\text{Find } (p^\varepsilon,q^\varepsilon,l^\varepsilon) \in \mathcal{G}\times\mathcal{V}\times \mathcal{G} \,, \text{ such that } \\
  &\left\{
  \begin{array}[c]{ll}
    (A_\perp \nabla_\perp p^\varepsilon, \nabla_\perp \eta) + (A_\perp \nabla_\perp q^\varepsilon, \nabla_\perp \eta)  = (f,\eta) \,, \, &\forall \eta \in \mathcal{G} \,, \\[2mm]
  (A_\parallel \nabla_\parallel q^\varepsilon, \nabla_\parallel \xi) + \varepsilon (A_\perp \nabla_\perp (q^\varepsilon+p^\varepsilon), \nabla_\perp \xi) + (l^\varepsilon,\xi) = (f,\xi) \,, \, &\forall \xi \in \mathcal{V} \,, \\[2mm]
(q^\varepsilon,\chi)=0 \,, & \forall \chi \in \mathcal{G} \,.
  \end{array}\right.    
  \end{split}
\end{equation*}
The fluctuations are functions of the non constrained space  $\mathcal{V}$, easily discretized  by standard numerical methods. This feature comes at the price of an additional unknown, namely $l^\varepsilon \in \mathcal{G}$ penalizing the property  $q^\varepsilon \in \mathcal{A}$.

The last difficulty is related to the discretization of the functional space $\mathcal{G}$ for the mean. To this end, the space $\mathcal{L}$ of Lagrange multipliers is introduced:
\begin{gather*}
  \mathcal{L} := \big\{ \lambda \in L^2(\Omega)\, | \,\nabla_\parallel \lambda \in L^2(\Omega), \lambda = 0 \text{ on } \Gamma_\text{in} \big\}\,, 
  \text{ with } \Gamma_\text{in} := \big\{ x\in \Gamma \, | \, b(x)\cdot n(x) < 0  \big\} 
\end{gather*}
such that the weak formulation of the problem can be stated as 
\begin{equation}
  \begin{split}
  &\text{Find } (p^\varepsilon,q^\varepsilon,l^\varepsilon,\lambda^\varepsilon,\mu^\varepsilon) \in \mathcal{V}\times\mathcal{V}\times \mathcal{V}\times\mathcal{L}\times\mathcal{L} \,, \text{ such that } \\
  &\hspace*{0em}\left\{
  \begin{array}[c]{ll}
    (A_\perp \nabla_\perp p^\varepsilon, \nabla_\perp \eta) + (A_\perp \nabla_\perp q^\varepsilon, \nabla_\perp \eta) + (A_\parallel \nabla_\parallel \eta,\nabla_\parallel \lambda^\varepsilon) \\
    \hspace*{7.2cm}= (f,\eta) \,, \, &\forall \eta \in \mathcal{V} \,, \\
    (A_\parallel \nabla_\parallel p^\varepsilon, \kappa) = 0\,,  & \forall \kappa \in \mathcal{L}\,,\\[2mm]
  (A_\parallel \nabla_\parallel q^\varepsilon, \nabla_\parallel \xi) + \varepsilon (A_\perp \nabla_\perp (q^\varepsilon+p^\varepsilon), \nabla_\perp \xi) + (l^\varepsilon,\xi)\\
    \hspace*{7.2cm}= (f,\xi) \,, \, &\forall \xi \in \mathcal{V} \,, \\
(q^\varepsilon,\chi)+(A_\parallel \nabla_\parallel \chi,\nabla_\parallel \mu^\varepsilon)=0 \,, & \forall \chi \in \mathcal{V} \,, \\
(A_\parallel \nabla_\parallel l^\varepsilon,\nabla_\parallel \tau) = 0\,, & \forall \tau \in \mathcal{L} \,.
  \end{array}\right.    
  \end{split}
\end{equation}
In this weak formulation of the singular perturbation problem, the discretization of the functional spaces $\mathcal{A}$ and $\mathcal{G}$ is not necessary. This is obtained thanks to the introduction of three auxiliary variables, namely  $l^\varepsilon$ penalizing the constraint $q^\varepsilon \in \mathcal{A}$, as well as  $\lambda^\varepsilon$  and $\mu^\varepsilon$ for the constraints $\eta \in \mathcal{G}$ and $\chi \in \mathcal{G}$. 

\medskip
Another route is proposed in \cite{degond_asymptotic-preserving_2012} operating a ``Micro-Macro'' decomposition of the solution. It consists in splitting $\phi^\varepsilon$ into two non orthogonal components:   $ \phi^\varepsilon = p^\varepsilon +\varepsilon \q^\varepsilon $, with $\q^\varepsilon \in \mathcal{L}$. The fluctuation space $\mathcal{A}$ is replaced by a space populated with functions vanishing on one part of the boundary. This space is readily discretized by standard numerical methods. On top of that, the problem is formulated for the unknowns $(\phi^\varepsilon,\q^\varepsilon)$, so that the discretization of the functional spaces is straightforward. The weak formulation of the problem is thus
\begin{equation*}
  \begin{split}
  &\text{Find } (\phi^\varepsilon,\q^\varepsilon) \in \mathcal{V}\times\mathcal{L}\,, \text{ such that }  \\
  &\left\{
  \begin{array}[c]{ll}
    (A_\perp \nabla_\perp \phi^\varepsilon, \nabla_\perp v) + (A_\parallel \nabla_\parallel \q^\varepsilon, \nabla_\parallel v) = (f,v) \,, \,\qquad  &\forall v \in \mathcal{V} \,, \\[2mm]
  (A_\parallel \nabla_\parallel \phi^\varepsilon, \nabla_\parallel w) = \varepsilon (A_\parallel \nabla_\parallel \q^\varepsilon , \nabla_\parallel w)    \,, \qquad &\forall w \in \mathcal{L} \,. \\
  \end{array}\right.    
  \end{split}
\end{equation*}
This system is Asymptotic-Preserving. Indeed, letting $\varepsilon \to 0$, the limit problem is recovered with 
\begin{equation*}
  \begin{split}
  &\text{Find } (\phi^0,\q^0) \in \mathcal{V}\times\mathcal{L}\,, \text{ such that }  \\
  &\left\{
  \begin{array}[c]{ll}
    (A_\perp \nabla_\perp \phi^0, \nabla_\perp v) + (A_\parallel \nabla_\parallel \q^0, \nabla_\parallel v) = (f,v) \,, \,\qquad  &\forall v \in \mathcal{V} \,, \\[2mm]
  (A_\parallel \nabla_\parallel \phi^0, \nabla_\parallel w) = 0    \,, \qquad &\forall w \in \mathcal{L} \,. \\
  \end{array}\right.    
  \end{split}
\end{equation*}
In this formulation  $\q^0$ is the Lagrangian associated to the constraint $\nabla_\parallel \phi^0 = 0$. 

\begin{table}[!ht]
  \centering
  \caption{Comparison of the Micro-Macro (M.-M.) and Duality Based (D. B.) methods with a FEM discretization of the singular perturbation problem (P) for coordinates non aligned to the anisotropy direction: number of rows, number of non zeros elements in the system matrix and computational time for the linear system resolution (thanks to the sparse direct solver MUMPS  \cite{amestoy_fully_2001,amestoy_hybrid_2006}) relative to that of the singular perturbation problem. The computations are carried out on a $100\times 100$ mesh with a  $\mathbb{Q}_2$ finite element method.}\label{tab:table3}

  \begin{tabular}{|c||c|r|r|}
    \hline\rule{0pt}{2.5ex}
    Meth. & \# rows &\multicolumn{1}{|c|}{\# Nnz} & \multicolumn{1}{|c|}{Time}\\
    \hline
    \hline\rule{0pt}{2.5ex}
    (M.-M.) &
    $20\times 10^{3}$ &
    $623\times 10^{3}$ &
    231\% 
    \\
    \hline\rule{0pt}{2.5ex}
    (D. B.) &
    $50\times 10^{3}$ &
    $1563\times 10^{3}$ &
    1478\% 
    \\
    \hline\rule{0pt}{2.5ex}
    (P) &
    $10\times 10^{3}$ &
    $156\times 10^{3}$ &
    100\% 
    \\
    \hline
  \end{tabular}
\end{table}
Both methods provide the same Asymptotic-Preserving properties with coordinates and meshes not related to the anisotropy. The efficiency of the Micro-Macro and the Duality based formulations is compared, in the table~\ref{tab:table3}, to a standard FEM discretization of the singular perturbation problem for two dimensional computations. The introduction of the three Lagrangians increases significantly both the system matrix size and the number of non zero elements. The Micro-Macro scheme is much more efficient with a matrix size only twice as big as that of the standard method and a computational time increasing roughly to  the same extent.

\bigskip

The Micro-Macro as well as the duality based methods have been developed to heterogeneous anisotropy ratios in \cite{degond_duality-based_2010,degond_asymptotic-preserving_2012}. The Micro-Macro approach has been extended to closed magnetic field lines thanks to a regularization of the problem introduced in \cite{narski_asymptotic_2013} and also implemented in \cite{deluzet_numerical_2015}. Note that, an asymptotic preserving method based on a lagrangian integration along the anisotropy direction is proposed in \cite{chacon_asymptotic-preserving_2014}.




\section{Conclusions}
In this document, Asymptotic-Preserving methods are reviewed in the frame of three singular perturbation problems. First the concept of Asymptotic-Preserving methods are outlined for the quasi-neutral limit of plasma descriptions. The scale of interest here is the Debye length measuring the typical size of the space charge creations. In this context, the derived AP-schemes offer the possibility to choose the model harnessed accordingly to the needs of the physics. Indeed, when the discretization parameters are large compared to the Debye length, AP-schemes define a consistent discretization of a quasi-neutral model, with properties similar to Magneto-Hydro-Dynamic descriptions (however with a finite electron inertia and eventually a kinetic description). Conversely, local upscaling are possible, by adjusting the mesh size to the local Debye length, the numerical methods becoming consistent with the non quasi-neutral model (either the Vlasov-Maxwell or the Euler-Maxwell system depending on the description used for the plasma, electrostatic models being also proposed).

The systematic methodolgy implemented to derive AP scheme is decomposed in different steps. First the limit regime needs to be clearly identified. It consists of a set of equations providing the limit of the solution in the asymptotic regime. In the framework of singular perturbation problem, this set of equations is not readily obtained from the initial multiscale problem. The asymptotic analysis is a crucial tool to clarify the inter relations of the multiscale and the limit problems.
More importantly, the derivation of the limit (quasi-neutral) problem from the multiscale (non quasi-neutral) set of equations is a key point in the construction of Asymptotic-Preserving methods. This preliminary work is thus capitalized on to manufacture a set of reformulated equations in which the quasi-neutral limit is regular: the limit problem is recovered by setting formally the asymptotic parameter to zero in the reformulated system. In this specific context, the asymptotic parameter is defined as the ratio of the Debye length and the typical length of interest, or equivalently the mesh size. Until this stage all the analysis are free from any numerical method. Finally, the question of the discretization is addressed. Performing the same analysis but with discretized equations, allows to derive an asymptotic preserving numerical method.

Two other frameworks are also addressed in the present document. The drift regime for fluid description of tokamak plasmas, also referred to as the gyro-fluid limit. The asymptotic regime investigated is the limit of infinite acoustic wave speeds (low Mach regime) with local transitions to flows characterized by Mach numbers close to one. In the low Mach regime, the parallel momentum equation degenerates into an equilibrium relation, with a pressure gradient balancing the electric force. No contribution of the parallel momentum appears explicitly in this equation anymore, which is at the origin of the singular nature of the drift limit. The applications envisioned here are more specifically related to the tokamak plasma edge physics, with a different dynamic in the plasma core compared to that of the sheath created in the vicinity of the wall. The derivation of asymptotic preserving schemes, in both the frameworks of the quasi-neutral and the drift regimes, involves the resolution of anisotropic elliptic or diffusion problems. This class of equations define a singular perturbation problem when the boundary conditions supplementing the diffusion operator in the aligned direction (with respect to the magnetic field) translate, for instance, the periodicity of the torus. The last section of this document is thus dedicated to a review of asymptotic preserving methods developed to address efficiently these problems.

The systematic derivation of AP-methods can be extended to more singular limits. Some examples can be named, with an extension of the quasi-neutral limit investigations aiming to bridge the Vlasov-Maxwell system and a Magneto-Hydrodynamic model or the Vlasov-Poisson system and the Boltzmann relation for the electrons. The adiabatic response as well as MHD models are successfully operated for the simulation, because they filter out from the equations most of the high frequencies and give access to the macroscopic evolution in an efficient way. In both cases the limit is much more singular, since in addition to the quasi-neutral asymptotic, a fluid and a mass-less (for the electron) limits are necessary to define the reduced models. Another extension can be considered in the gyro-fluid framework, with the investigation of the quasi-neutrality break down, in order to address extensively the regimes transition between the plasma core and the sheath. 
A more distant issue can also be envisioned, with the extension of AP-methods to other scientific fields where multiscale problems are common, namely biology and complex systems for instance, where this class of methods, if already investigated, lack significant developments.

\section*{Acknowledgments} This work has been carried out within the framework of the EUROfusion Consortium and has received funding from the Euratom research and training programme 2014-2018 under grant agreement No 633053. The views and opinions expressed herein do not necessarily reflect those of the European Commission.\\
Furthermore, FD would like to acknowledge support from the ANR MOONRISE (MOdels, Oscillations and NumeRIcal SchEmes, 2015-2019) and partial support from ANR PEPPSI (Plasma Edge Physics and Plasma-Surface Interactions, 2013-2017).
This work has been supported by the National Science Foundation (NSF) under
grant RNMS11-07444 (KI-Net). PD is on leave from CNRS, Institut de
Math\'ematiques de Toulouse, France. PD acknowledges support from the Royal
Society and the Wolfson foundation through a Royal Society Wolfson Research
Merit Award.
\bibliographystyle{abbrv}
\bibliography{bib}


\end{document}